\documentclass[fleqn,usenatbib]{mnras}

\usepackage{newtxtext}
\usepackage[varvw]{newtxmath}
\usepackage{hyperref}
\usepackage[T1]{fontenc}

\DeclareRobustCommand{\VAN}[3]{#2}
\let\VANthebibliography\thebibliography
\def\thebibliography{\DeclareRobustCommand{\VAN}[3]{##3}\VANthebibliography}

\usepackage{graphicx}	% Including figure files
\usepackage{amsmath}	% Advanced maths commands
\usepackage{tabularx}
\title[Blazar variability with \textit{Swift}-BAT]{Long-term hard X-ray variability properties of \textit{Swift}-BAT blazars}

\author[Mundo \& Mushotzky]{
Sergio A. Mundo$^{1}$\thanks{E-mail: smundo@astro.umd.edu}
and Richard Mushotzky
\\
$^{1}$Department of Astronomy, University of Maryland, College Park, MD 20742, USA\\
}

\date{Accepted XXX. Received YYY; in original form ZZZ}

\pubyear{2023}

\begin{document}
\label{firstpage}
\pagerange{\pageref{firstpage}--\pageref{lastpage}}
\maketitle

% Abstract of the paper
\begin{abstract}
We present results from the first dedicated study in the time domain of the hard X-ray variability behavior of blazars on long timescales based on $\sim$13 years of continuous hard X-ray data in the 14-195 keV band. We use monthly-binned data from the recent 157-month \textit{Swift}-BAT catalog to characterize the hard X-ray variability of 127 blazars and search for potential differences between the variability of BL Lacertae objects (BL Lacs) and flat-spectrum radio quasars (FSRQs). A significant portion of the blazars in the sample ($\sim$37\%) do not show statistically significant hard X-ray variability on monthly timescales, which is deeply at odds with previous studies that show that blazars are highly variable in the X-rays and other energy bands on a wide range of timescales. We also find that, on average, the FSRQs and BL Lacs for which we do detect variability exhibit similar flux variability; this suggests that the variability in these FSRQs is not necessarily driven by variations in the source function of scattered external radiation arriving from extended regions, and that it is instead possibly driven by processes that lead to variations in particle injection. In addition, only five blazars in our sample show significant spectral variability in the long-term light curves. For 3 blazars, we find that a power law that changes slope on monthly timescales is sufficient to characterize the variable hard X-ray spectrum, suggesting that, at least for some bright blazars, the long-term spectra in the hard X-rays may be described in a relatively simple fashion.  

%($\langle F_{\rm var} \rangle = 76\pm5$\% vs $\langle F_{\rm var} \rangle = 75\pm4$\%)

%the variable hard X-ray spectrum can be characterized by a power law that changes slope on monthly timescales, suggesting that, at least for some of the brightest blazars, the long-term spectra in the hard X-rays may be described in a relatively simple fashion.                
\end{abstract}

% Select between one and six entries from the list of approved keywords.
% Don't make up new ones.

%%%%%%%%%%%%%%%%%%%%%%%%%%%%%%%%%%%%%%%%%%%%%%%%%%

%%%%%%%%%%%%%%%%% BODY OF PAPER %%%%%%%%%%%%%%%%%%

\section{Introduction}

Active galactic nuclei (AGN) are known to exhibit intrinsic stochastic variability on a wide range of timescales, and the study of such variability provides insight into the physical processes and structures associated with the emitting source. Blazars are radio-loud (RL) AGN characterized by having a jet pointed close to the line of sight of the observer. The particles in the jet move with bulk relativistic speed $v = \beta c$ towards the observer, and as a result emission originating in the jet is relativistically boosted in the direction of motion. More specifically, an observer at rest will detect a broadband flux that scales as $F = \delta^{4}F'$ compared to the emission in the co-moving frame of the jet, where $\delta = [\Gamma(1-\beta \rm cos \theta)]^{-1}$ is defined as the Doppler factor, and $\Gamma$ and $\theta$ are the bulk Lorentz factor and viewing angle, respectively (\citealt{1995PASP..107..803U}). These conditions therefore lead to emission in the observer's frame that is much more powerful than if the jetted particles were at rest; consequently, over the past several decades, the mostly non-thermal emission from blazars has been characterized by high luminosities, as well as by high-amplitude, rapid variations in flux, spectra, and polarization across most timescales and wavelengths \citep[see e.g.][]{1976ARA&A..14..173S,1978PhyS...17..265B,1980ARA&A..18..321A,1985ApJ...298..114M,1986ApJ...306L..71M,1986ApJ...302..337F,1995ARA&A..33..163W,1997ARA&A..35..445U,2005A&A...442...97A,2008A&A...486..721L,2014A&A...563A..57S}.

In general, blazars can be classified into two categories: BL Lacertae (BL Lac) objects and flat-spectrum radio quasars (FSRQs). Historically, the distinction between these two types has been based on the rest-frame equivalent width (EW) of optical emission lines: FSRQs exhibit broad emission lines in their optical spectra (EW$>$5\AA), while BL Lacs have featureless optical spectra or spectra with weak emission lines \citep[]{1991ApJ...374..431S}. The two subclasses can also be distinguished via their spectral energy distributions (SEDs). The SEDs of all blazars generally consist of a double-hump structure, with the hump at lower frequencies caused by synchrotron emission from the radio to the ultraviolet (UV)/X-rays and the hump at higher frequencies (X-rays to $\gamma$-rays) arising due to inverse Compton processes. In BL Lacs, the two humps are usually nearly equally luminous, and the Compton hump is likely due to the synchrotron self-Compton (SSC) mechanism, where the synchrotron photon field produced by the jet is Compton upscattered by the same highly energetic particles in the jet. For FSRQs, the inverse Compton hump is often significantly more luminous than the synchrotron hump, and is instead believed to be the product of inverse Compton scattering of photons external to the jet (the so-called ``external Compton" (EC) process). This has led to a more physical divide between BL Lacs and FSRQs, where the more luminous FSRQs likely exhibit radiatively efficient accretion; this forms a UV-bright disk that photo-ionizes the broad-line region (BLR), which in turn provides the external photon field that is upscattered by the jet \citep[]{2009MNRAS.396L.105G,2011MNRAS.414.2674G}. FSRQs and BL Lacs can further be divided based on the frequency of the peak of the synchrotron hump, $\nu_{\rm syn}^{\rm peak}$, leading to the terms ``high-synchrotron-peaked" (HSP, $\nu_{\rm syn}^{\rm peak} \gtrsim 10^{15}$Hz); ``intermediate-synchrotron-peaked" (ISP, $10^{14} \lesssim \nu_{\rm syn}^{\rm peak} \lesssim 10^{15}$Hz); and ``low-synchrotron-peaked" (LSP, $\nu_{\rm syn}^{\rm peak} \lesssim 10^{14}$Hz; \citealt{2010ApJ...716...30A}), with FSRQs usually classified as LSPs and the ``blazar sequence" dictating that the bolometric luminosity gradually decreases as one goes from FSRQ/LSP to HSP \citep[]{1998MNRAS.299..433F,2017MNRAS.469..255G}.

When it comes to the production of the X-rays in particular, the emission mechanisms involved in BL Lacs and FSRQs can naturally lead to differences in the nature of the variability for each type of blazar. For instance, the X-rays in BL Lacs are usually produced by either synchrotron emission or the SSC process (see e.g. \citealt{2009MNRAS.396L.105G}). In this case, the blazar's variability is likely driven by the timescale for the acceleration of particles in the jet as it compares to the timescale of energetic losses by those same particles through either synchrotron or inverse Compton processes. By contrast, as implied earlier, the X-rays in FSRQs are produced by the EC process, where the dominant radiation field corresponds to the photon density of an external component. The variability in these blazars could involve variations in an external photon field that scatters off a quasi-stable particle distribution. The implications that the aforementioned processes have on the characterization of the X-ray variability of blazar subclasses remain a major point of discussion, as differences in variability between different spectral components and between blazar types may provide insight into the physical mechanisms involved in driving the jetted emission.

%The implications that the aforementioned processes have on the characterization of the variability of blazar subclasses remain a major point of discussion, as differences in variability between different spectral components and between blazar types may provide insight into the physical mechanisms involved in driving the jetted emission.

%The implications that the aforementioned processes have on the characterization of the variability of BL Lacs and FSRQs remain a major point of discussion, as differences in variability between different spectral components and between blazar types may provide insight into the physical mechanisms involved in driving jetted emission.

The X-ray portion of blazar SEDs can generally be well-modeled by a simple power law representing the non-thermal processes occurring in the jet (e.g. \citealt{1996ApJ...463..424U,2000ApJ...533..650S,2019ApJ...881..154P}), although curvature in the form of a log-parabola or a broken power law has also been detected in a significant number of objects \citep[e.g.][]{1997ApJ...480..534C,1998ApJ...504..693K,2004A&A...413..489M,2006A&A...448..861M,2009A&A...504..821P,2013ApJ...770..109F,2018A&A...616A.170A,2021MNRAS.508.1701D}. According to many blazar emission models, these curved continua may result from relativistic particle distributions that have a similar shape (see e.g. \citealt{1994ApJ...421..153S,1997ApJ...484..108S,2009ApJ...704...38S,2007ApJ...665..980T,2008MNRAS.386..945T,2009MNRAS.397..985G,2015MNRAS.448.1060G,2018A&A...616A.170A}). In addition, some sources have shown a ``harder-when-brighter" trend (i.e. the spectrum flattens as the source brightens, e.g. \citealt{2006ApJ...651..782Z,2014A&A...563A..57S,2015ApJ...807...79H,2017ApJ...841..123P,2018A&A...619A..93B}), possibly due to an increased contribution of a hard tail produced by the jet when the source is in a brighter state.  The behavior of blazar X-ray spectra is therefore expected to be fairly variable.

%increased particle injection that hardens the particle energy distribution

%Other studies have instead a ``softer-when-brighter" trend in some sources (see e.g. Hayashida et al. 2015)

%When it comes to the production of X-rays in particular, the emission mechanisms involved in BL Lacs and FSRQs can naturally lead to differences in the nature of the variability for each type of blazar. For instance, the X-rays in BL Lacs are usually produced by either synchrotron emission or the SSC process. In this case, the blazar's variability is likely driven by the timescale for the acceleration of particles in the jet as it compares to the timescale of energetic losses by those same particles through either synchrotron or inverse Compton processes. By contrast, as implied earlier, the X-rays in FSRQs are produced by the EC process, where the dominant radiation field corresponds to the photon density of an external component. The variability in these blazars therefore involves variations in an external photon field that scatters off a quasi-stable particle distribution. The implications that the aforementioned processes have on the characterization of the variability of blazar subclasses remain a major point of discussion, as differences in variability between different spectral components and between blazar types may provide insight into the physical mechanisms involved in driving the jetted emission.   

Blazar variability analyses in the X-rays have been instrumental in helping constrain the jet physics involved in these sources, both in multi-wavelength and X-ray-focused studies (see e.g. \citealt{1990ApJ...360..396W,1995PASP..107..803U,1996ApJ...463..424U,1996ApJ...470L..89T,1997ARA&A..35..445U,2000ApJ...533..650S,2002PASA...19...49P,2005ApJ...629..686Z,2006ApJ...651..782Z}, and references therein). While most of these analyses focused on X-ray observations taken within the ``classical" 0.3-10 keV range, recently, the hard X-rays ($\gtrsim10$ keV) in blazars and beamed AGN have been studied more extensively than in the past (e.g. \citealt{2015ApJ...812...14M,2015ApJ...807...79H,2017MNRAS.466.3309R,2018A&A...619A..93B,2022ApJ...932..104R}). The origin of the hard X-ray emission discussed in these studies may provide insight into the physics occurring in the innermost regions located near the base of the jet, with some sources serving as potential laboratories to probe the elusive disk-jet connection (\citealt{2016MNRAS.462.1542S,2018ApJ...859L..21C}). However, most of the work on blazars in the X-ray band has historically been heavily biased towards bright sources and/or sources in bright/active states; in the hard X-rays in particular, studies have also been limited to relatively shorter timescales (due to the observing strategies of the available hard X-ray telescopes and their limited field of view), thus leaving an incomplete picture of the X-ray variability behavior of blazars.

The Burst Alert Telescope aboard the \textit{Swift} observatory (\textit{Swift}-BAT) has a very wide field of view ($\sim60^{\circ}\times100^{\circ}$), and its main objective is to detect transient gamma-ray bursts (GRBs) with a coded-aperture mask telescope. However, as it searches for GRBs and other transients in the hard X-rays, it also continuously observes the sky, performing an all-sky hard X-ray survey in the 14-195 keV band (\citealt{2008ApJ...681..113T,2010ApJS..186..378T}). The recent release of the 105-month catalog (\citealt{2018ApJS..235....4O}), as well as the pending publication of the 157-month survey\footnote{https://swift.gsfc.nasa.gov/results/bs157mon/} (Lien et al. in prep), are unique in that they provide continuous, well-sampled observations over a $\gtrsim9$-year timescale for a hard X-ray selected sample of sources, and thus sample the time variability of these objects in a previously unexamined time domain. Due to the wide field of view and moderate sensitivity, this survey yields data for over 1600 sources that have been detected since $\sim$December 2004, $\sim$1000 of which are AGN, and 158 of which are classified as ``beamed AGN". The BAT catalog therefore currently boasts the largest sample of AGN observed in the hard X-rays on long timescales, with a wide range in redshift and luminosity exhibited across the AGN listed in the catalog.

In this paper, we aim to remove some of the bias towards bright blazars seen in past X-ray studies by studying 127 blazars from the 157-month BAT catalog (regardless of their brightness) and performing the first dedicated study of the hard X-ray variability of blazars on long timescales based on $\sim$13 years of \textit{Swift}-BAT data. The structure of the paper is as follows: we describe our sample selection and the BAT data and their filtering in Section \ref{sec:obs}, present our variability analysis and its results in Section \ref{sec:results}, and discuss our results in Section \ref{sec:sec4}.

\section{\textit{Swift}-BAT sample selection and data filtering} \label{sec:obs}

In the more recent \textit{Swift}-BAT catalogs, the ``QSO" type is replaced with either ``Seyfert" or ``beamed AGN" depending on the properties of the optical emission lines in the literature, and the Roma blazar catalog BZCAT (\citealt{2009A&A...495..691M}) is also used as a reference to classify beamed AGN (see e.g. \citealt{2018ApJS..235....4O}). This classification yields 158 sources that are defined as ``beamed AGN" in the BAT 157-month catalog. However, not all of these sources show the spectral shape that is expected of a blazar in their SEDs. As in \cite{2019ApJ...881..154P}, we exclude sources for which this is the case and therefore end up with a sample of 127 blazars for our analysis.

As for previous catalogs (see e.g. \citealt{2010ApJS..186..378T,2013ApJS..207...19B}), the 157-month catalog data reduction is carried out by extracting data in eight energy bands (14–20 keV, 20–24 keV, 24–35 keV, 35–50 keV, 50–75 keV, 75–100 keV, 100–150 keV, and 150–195 keV) from a single snapshot image; the data from each snapshot are then combined into all-sky mosaic images, which in turn are combined to form a total-band map, and a blind search for sources in the 14-195 keV band images is performed with a 4.8$\sigma$ detection threshold. Monthly- binned light curves are generated by creating monthly, all-sky total-band mosaic images, and then extracting the mosaic fluxes for each month for all sources detected in the full survey. The catalog includes monthly light curves for the total 14-195 keV BAT band, as well as light curves divided by the eight BAT bands; these are the data we use to conduct our long-timescale variability analyses.

\begin{figure}
 	% To include a figure from a file named example.*
 	% Allowable file formats are eps or ps if compiling using latex
 	% or pdf, png, jpg if compiling using pdflatex
 	\includegraphics[width=\columnwidth]{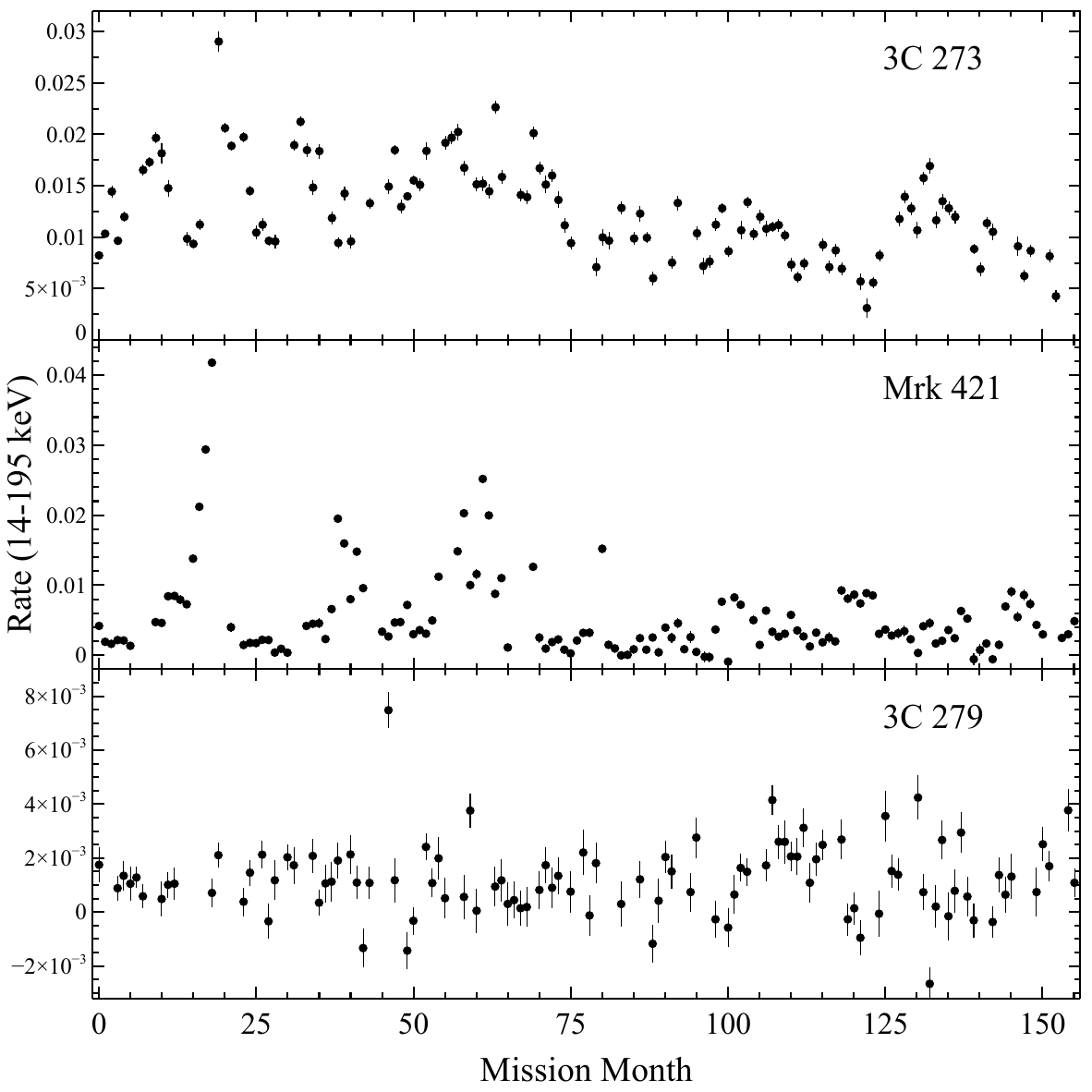}
     \caption{Examples of filtered monthly-binned light curves for 3 well-known blazars from the 157-month catalog in the 14-195 keV band. We exclude points for which the exposure is $<$1 day. Mission month 0 corresponds to the first month of the \textit{Swift}-BAT survey observations, namely December 2004 (see e.g. \citealt{2018ApJS..235....4O}).}
     \label{fig:157lcs}
\end{figure}

As in \cite{2014A&A...563A..57S}, we filter the light curves by excluding any data points with exposure times shorter than one day. In order to filter out points with very large error bars, we also inspect the histogram of the flux uncertainties of all light curves in logarithmic space to
determine where the high-value tail of the distribution begins. However, we find that for the most part, each of these two flags accounts for the same data points, so we use the flag on the exposure time as our main filter for the BAT monthly data (see Figure \ref{fig:157lcs} for examples of the filtered light curves).

\section{Results} \label{sec:results}

\subsection{Flux variability analysis in the 14-195 keV band}

We initially estimate the variability in our sources by fitting the 14-195 kev total band monthly light curves with a constant function, and then applying a $\chi^{2}$ test. The criterion that we set to represent a significant detection of the variability is $p_{\chi^{2}} < 5$\%, where $p_{\chi^{2}}$ is the null-hypothesis probability of obtaining that value of $\chi^{2}$ if the source were in fact constant. We also quantify the flux variability of our sources by following the methods described in e.g. \cite{2003MNRAS.345.1271V}, where the contribution of an additional variance from measurement uncertainties is corrected for. The “excess variance” has been widely used in studies of the variability of accreting objects to estimate the intrinsic source variance \citep[e.g.][]{1997ApJ...476...70N,2002ApJ...568..610E}, and is frequently normalized to directly compare the variance between different sources. The fractional root mean square (rms) variability amplitude $F_{\rm var}$ is defined as 

%defined as the square root of the normalized excess variance \textcolor{blue}{(see Eq. \ref{eqn:fvar})}, is the quantity we will use here.

\begin{equation}
\label{eqn:fvar}
    F_{\rm var} = \sqrt{\frac{S^{2} - \overline{\sigma_{\rm err}^{2}}}{\overline{x}^{2}}}, 
\end{equation}

\noindent i.e. the square root of the normalized excess variance, and has the added benefit of being a linear statistic, thus allowing for the representation of the rms variability amplitude in percentage terms; this is the quantity we will use here.

Upon applying the above methodology, we find that 6 sources exhibit a fractional variability that ends up being much higher than expected ($\gtrsim$240\%) upon inspection of their light curves. We believe that, for these sources, the high value of $F_{\rm var}$ is due to the fact that $\gtrsim$40\% of the data points correspond to negative count rates, implying a mean flux that is very close to 0; when we inspect their light curves, we do not see significant variability in the amplitudes, especially when taking into account their uncertainties. We therefore exclude these 6 sources from the rest of the analysis due to their relatively low signal-to-noise (S/N) data (for a list of the remaining 121 sources, see Table \ref{tab:blazarprops}).

We also find that a significant fraction ($\sim$37\%) of our sample of blazars does not show statistically significant variability on monthly timescales. For the vast majority of the objects in this sub-sample, we cannot measure $F_{\rm var}$ because we calculate a slightly negative normalized excess variance (i.e., the uncertainties are slightly larger than the sample variance), suggesting that any variation in the amplitudes is fairly low. This is surprising, since it has been established from past studies that the emission from blazars is extremely variable at almost every timescale and wavelength. It is important to note, however, that the BAT has a relatively low sensitivity per unit time (\citealt{2008ApJ...681..113T}), and that it is not clear from the BAT data alone if the lack of variability we observe is actually a result of relatively constant emission. In a joint analysis with \textit{NICER} data (\citealt{2023MNRAS.520.1044M}), we show that, for at least 4 of the sources in this sub-sample, there is in fact detectable variability on monthly timescales and shorter, but that it is significantly lower-amplitude than what is expected of blazars (see Sec. \ref{sec:sec4} for further discussion).

\begin{figure}
	% To include a figure from a file named example.*
	% Allowable file formats are eps or ps if compiling using latex
	% or pdf, png, jpg if compiling using pdflatex
	\includegraphics[width=\columnwidth]{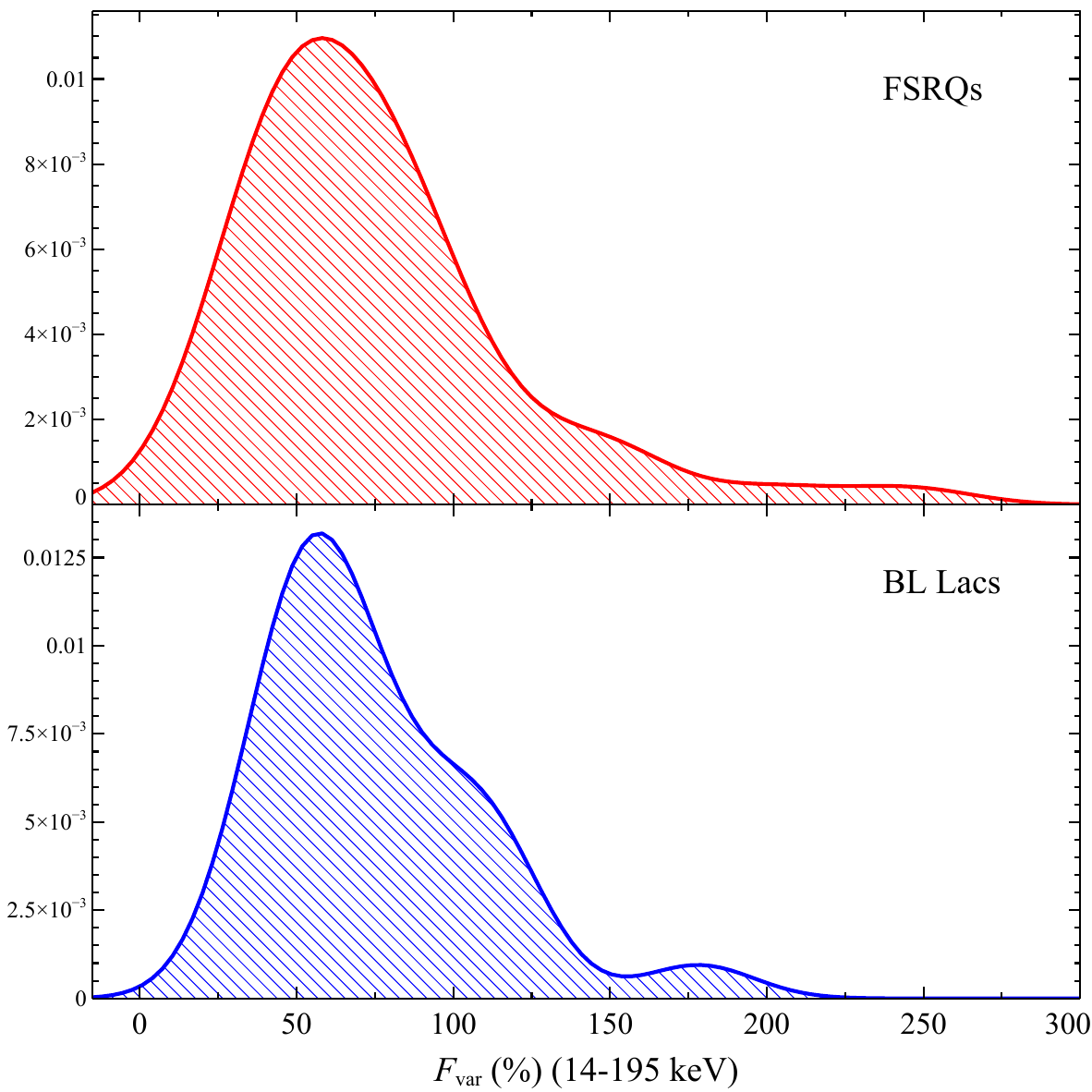}
    \caption{KDE distributions for $F_{\rm var}$. A KS-test shows that the two distributions are likely not significantly different, i.e. they may be drawn from the same parent population.}
    \label{fig:fvardist}
\end{figure}

We divide the remaining blazars for which we can detect statistically significant variability (76 total) into BL Lacs and FSRQs based on the classification in the BZCAT catalog (\citealt{2015Ap&SS.357...75M}), as well as on the SEDs in \cite{2019ApJ...881..154P}. We calculate $F_{\rm var}$ for each population (we use this sample for the rest of the analysis). The $F_{\rm var}$ distributions for each blazar type are shown in Figure \ref{fig:fvardist}, produced by using a kernel density estimate (KDE) of their probability density functions\footnote{In this paper, we visualize our distributions with KDEs. Each data point in the sample is assigned a specific kernel, and the corresponding densities are then summed together. Here we choose a Gaussian kernel with bandwidth corresponding to ``Scott's Rule" (\citealt{1992mde..book.....S}), i.e. width $=N^{-1/5}$ ($N$ is the number of data points). The benefit of using KDEs over regular histograms is that in the latter, the appearance can change based on the binning properties; a KDE only depends on the kernel width.}. We find that, within error bars, the variability between the two blazar types is practically the same, with mean $F_{\rm var}$ values of 76$\pm$5\% and 75$\pm$4\% for FSRQs and BL Lacs, respectively. Given that the distributions are fairly skewed, we also report the median values to compare the distributions, and find that these are identical (both 65\%). A Kolmogorov-Smirnov (KS) test confirms that the two distributions are likely not significantly different at the 95\% confidence level (KS statistic = 0.18, $p$ value = 0.60). Since, for most FSRQs, the X-ray emission is produced via the EC mechanism, one might expect an overall lower variability compared to BL Lacs; the reason for this is that the variability would be caused by variations in the scattered external radiation field, which arrives from extended regions such as the BLR and dusty torus, and would thus involve longer timescales and less rapid variations. Our results suggest that it is possible that the variability in FSRQs may instead be driven by variations in the particle energy distribution and particle injection processes.

\begin{table*}
	\centering
	\caption{Sample of 121 blazars used in this study, along with associated properties. 76 of these blazars are used in the correlation analysis between $F_{\rm var}$ and the last 4 parameters in this table. (1) is the blazar name; (2) is the blazar type; (3) is the source redshift; (4) is the luminosity in the BAT band; (5) is the black hole mass estimate, gathered from the literature if available; (6) is the Doppler factor (from \citealt{2019ApJ...881..154P}); (7) is the spectral index for the time-averaged spectra in the BAT band; (8) is the fractional variability. Empty entries are meant to indicate that the parameters were not readily found in the literature. Sources for which we do not detect statistically significant variability have a dash in column (8). The full list of sources can be found in the electronic version and in the 157-month BAT catalog.}
	\label{tab:blazarprops}
    \resizebox{\textwidth}{!}{\begin{tabular}{lccccccc}
	%\begin{tabularx}{\textwidth}{lccccccc} % four columns, alignment for each
		\hline
		Name & Blazar Type & $z$ & $L_{\rm 14-195 \ keV}$ & $M_{\rm bh}$ & $\delta$ & $\Gamma_{\rm 14-195 \ keV}$ & $F_{\rm var}$ (\%)\\
		\hline
        (1) & (2) & (3) & (4) & (5) & (6) & (7) &\\
        \hline
		Mrk 1501 & FSRQ & 0.0893 & 44.78 & 8.70 & 9.5 & 1.82$^{+0.22}_{-0.21}$ & 59$\pm$9\\
		1ES 0033+595 & BL Lac & 0.0860 & 44.68 & 7.25 & 13.6 & 2.81$^{+0.22}_{-0.19}$ & 111$\pm$2\\
		PKS 0101-649 & FSRQ & 0.1630 & 45.02 & 8.70 & 18.2 & 1.58$^{+0.54}_{-0.51}$ & ---\\
		SHBL J012308.7+342049 & BL Lac & 0.2720 & 45.41 &  & 15.7 & 2.94$^{+0.61}_{-0.48}$ & ---\\
        B2 0138+39B & BL Lac & 0.0800 & 44.37 &  &  & 2.08$^{+0.44}_{-0.38}$ & 55$\pm$25\\
        ... & ... & ... & ... & ... & ... & ... & ...\\
		\hline
	\end{tabular}}
\end{table*}

\subsection{Correlation analysis} \label{subsec:spec}

In order to test for possible trends between the hard X-ray variability and some of the main properties of the 76 variable blazars in our sample, we correlate the fractional variability with the luminosity in the BAT band, black hole mass, Doppler factor, and the photon index $\Gamma$ from the time-averaged spectra in the 14-195 keV band, for the BL Lac and FSRQ populations separately. We use the FSRQ black hole mass estimates and the Doppler factors for both populations from \cite{2019ApJ...881..154P}, as well as the BL Lac masses estimated via the properties of the host galaxy (e.g. stellar velocity dispersion, fundamental plane relation) in studies such as \cite{2002ApJ...579..530W} and \cite{2005ApJ...631..762W}; we list the black hole masses of the blazars in column (5) of Table \ref{tab:blazarprops}. The masses of $\sim$9 BL Lacs in this sample are not readily available in the literature. We apply a Spearman rank correlation test to quantify any potential trends.  

\begin{figure*}
\centering
    \includegraphics[width=0.47\textwidth]{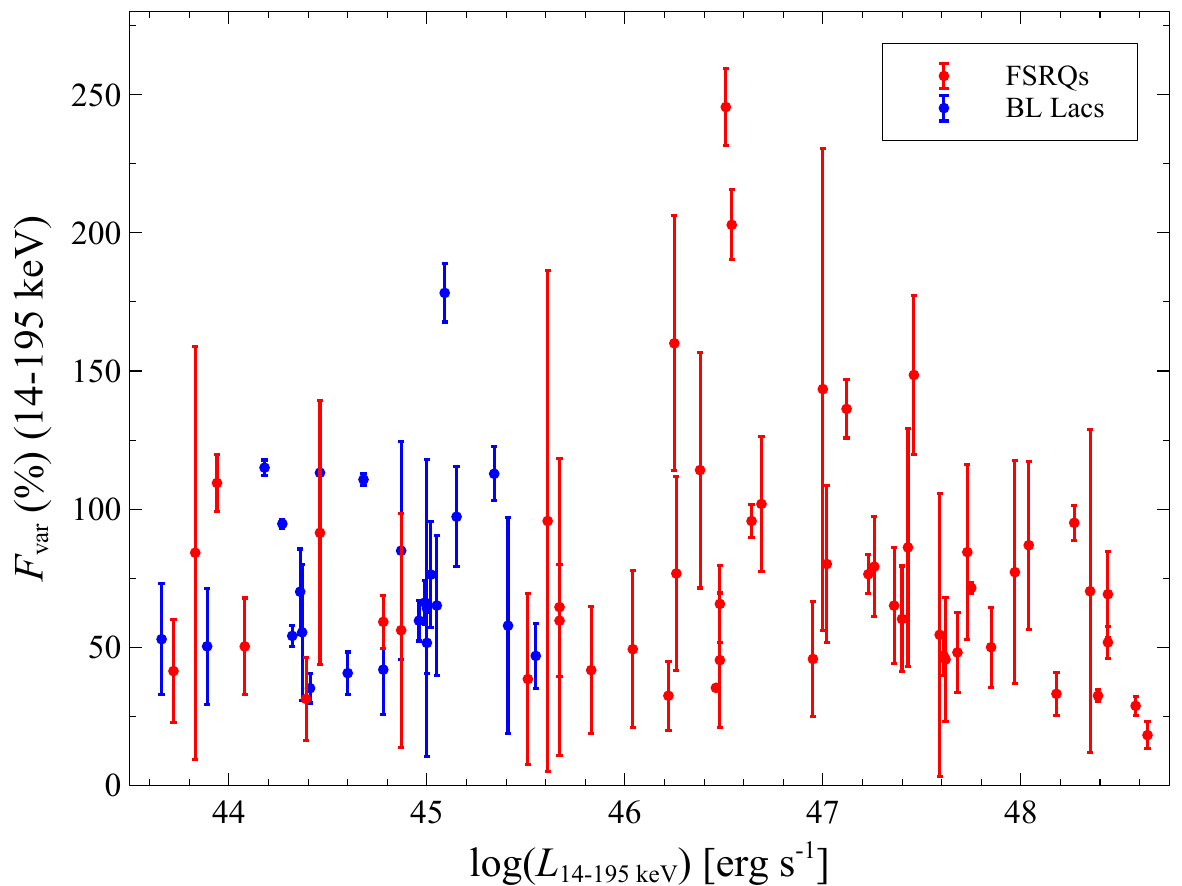}
    \includegraphics[width=0.47\textwidth]{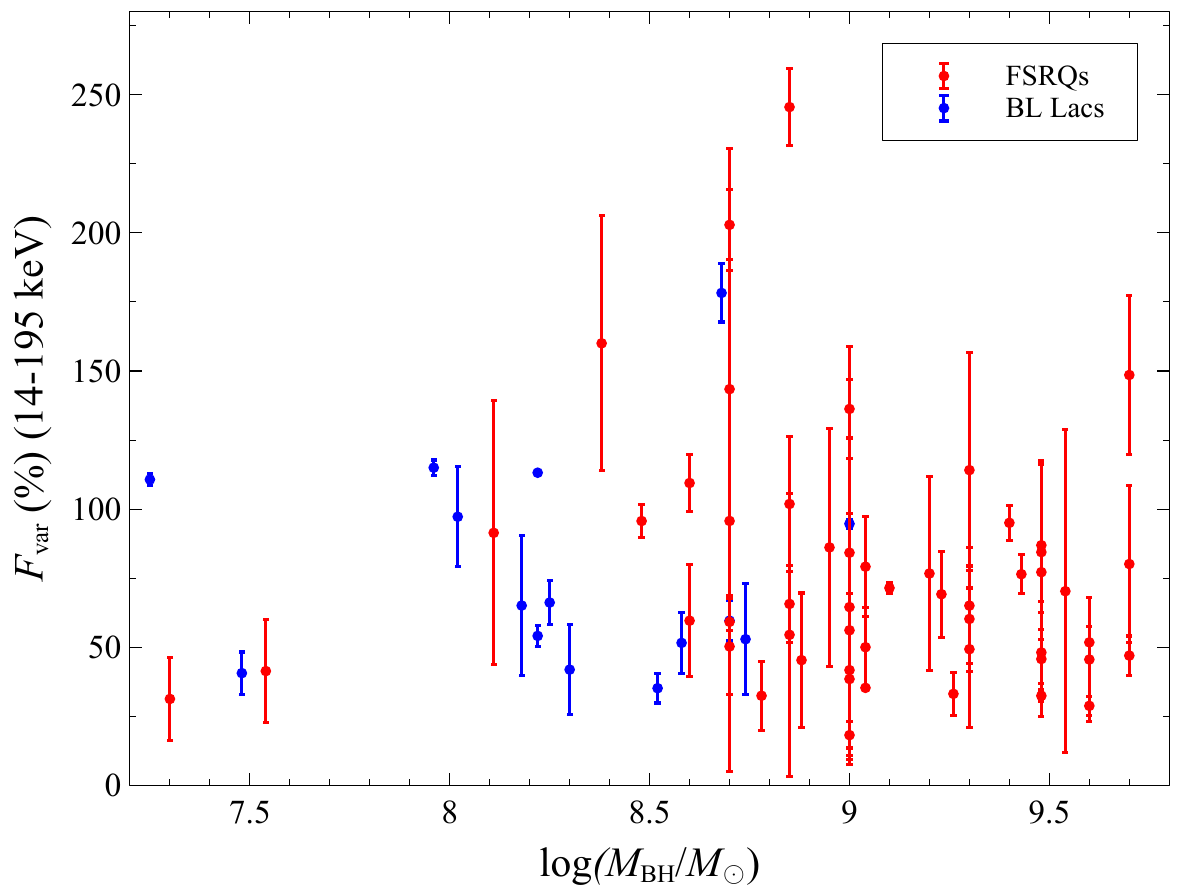}
    \includegraphics[width=0.47\textwidth]{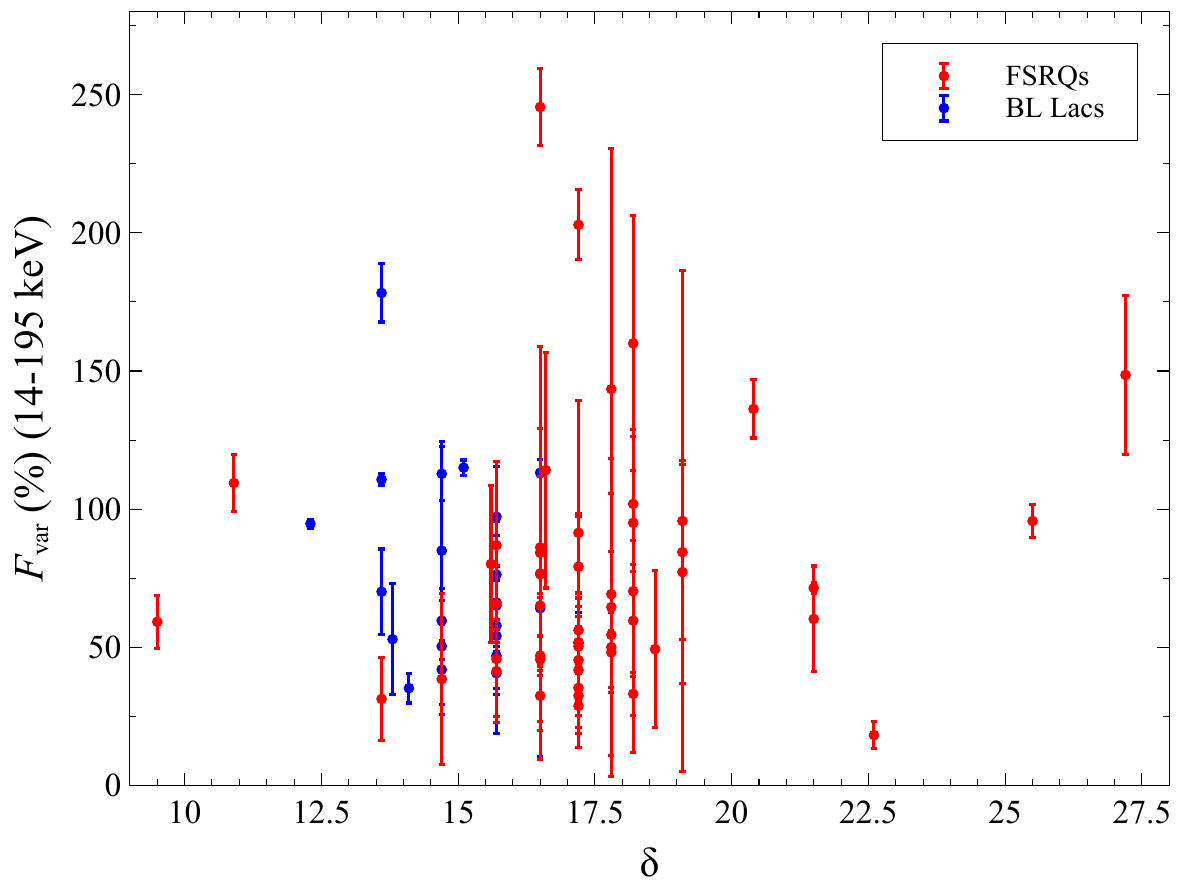}
    \includegraphics[width=0.47\textwidth]{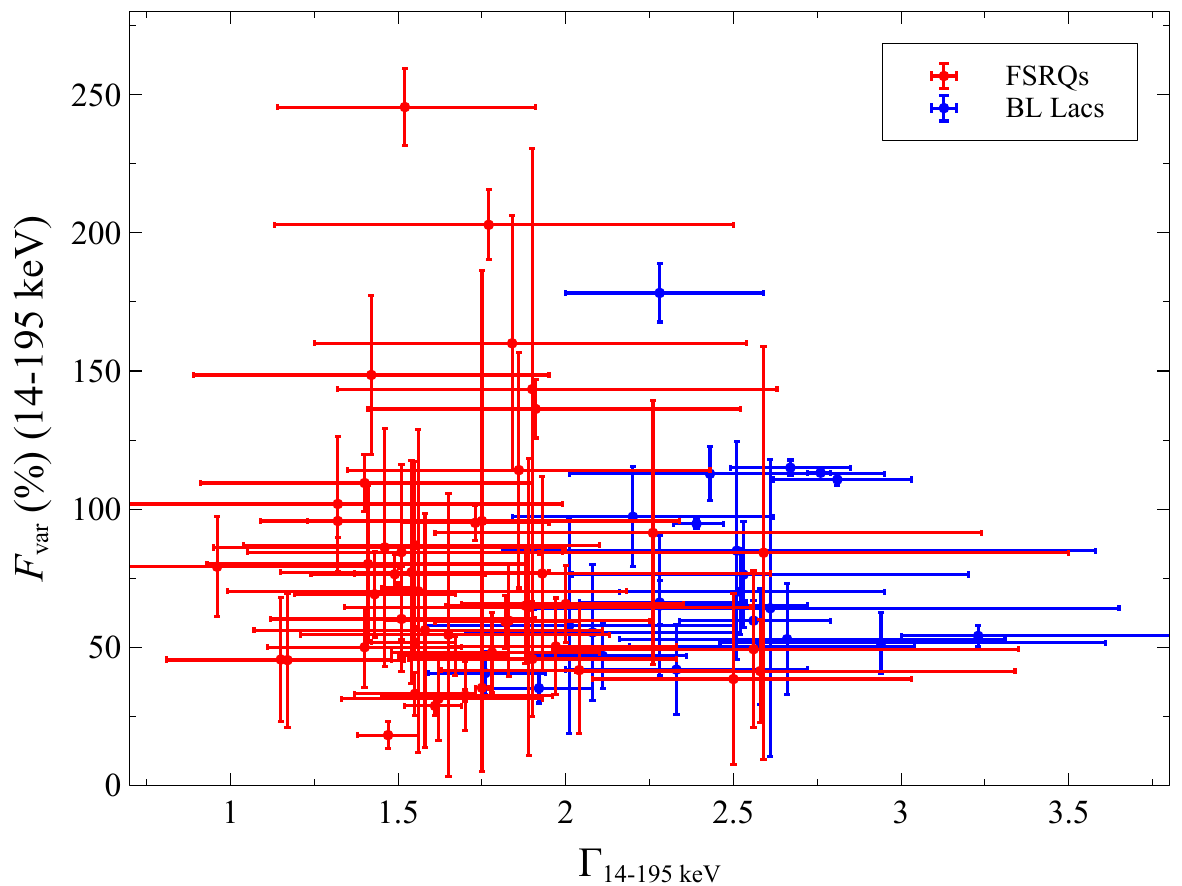}
\caption{$F_{\rm var}$ as a function of the BAT luminosity, black hole mass, Doppler factor, and photon index in the BAT band. We do not find statistically significant correlations between the hard X-ray variability and parameters that represent the properties of the blazars in our sample. (\textit{Note: The fairly high variability ($F_{\rm var} \gtrsim$175\%) seen in 3 sources in these plots is not due to low S/N, as was essentially the case for the 5 objects we previously excluded from this analysis.})}
\label{fig:corrs}
\end{figure*}

We plot $F_{\rm var}$ against each of these parameters in the 4 panels in Figure \ref{fig:corrs}. While we do not detect any significant correlation between the variability and these parameters for either blazar type (i.e., the probability that a correlation occurs by chance is always $>$ 0.05), the results still provide some potential insights. For instance, the lack of correlation between the variability and both the luminosity and black hole mass is in contrast to the anti-correlations that have been found in past studies within the 2-10 keV band for line-emitting AGN (Seyferts and quasars; see e.g. \citealt{1986Natur.320..421B,1993MNRAS.265..664G,1995MNRAS.273..923P,2004MNRAS.348..207P,2010ApJ...710...16Z,2011ApJ...730...52K,2013ApJ...779..187K,2012A&A...542A..83P}). However, it is important to note that these anti-correlations were usually observed in analyses that involved timescales shorter than those probed by the BAT, and therefore would likely not be detected in this study (see e.g. \citealt{2004ApJ...617..939M,2011nlsg.confE...8S,2011ApJ...726...21Z,2014A&A...563A..57S} for similar results on longer timescales). In addition, and probably more importantly, the aforementioned studies focused on Seyferts, which of course have very different underlying astrophysical processes associated with their X-ray emission (e.g. disk/coronal emission), and so the trends observed with these sources were usually directly related to the accretion process. It is therefore unclear what should have been expected to begin with when it came to our sample of blazars and their jetted emission (see Sec. 4 for further discussion).

%our plots confirm the contrast in luminosities between the two populations (Fossati et al. 1998, Ghisellini et al. 2017, Paliya et al. 2019), with FSRQs on average being significantly more luminous than BL Lacs. In addition,

Small changes in jetted emission will be Doppler boosted towards the observer to larger-amplitude variations; one might therefore expect to detect a positive correlation between the Doppler factor $\delta$ and the fractional variability. While we do not find such a correlation in our analysis, we do calculate a mean Doppler factor for the FSRQs in our sample that is larger than that for the BL Lacs ($\delta \sim 18$ vs $\delta \sim 15$, respectively). A KS-test shows that the Doppler factor distributions for the two blazar types are significantly different from each other (KS statistic = 0.70, $p$ value $\ll$ 0.05). This suggests that effects related to relativistic beaming may be a significant driving factor for the X-ray variability in these FSRQs, enhancing it to a level detectable by the BAT. 

%Given the higher than expected mean $F_{\rm var}$ that we find for FSRQs

%This is to be expected of the more luminous FSRQs, whose luminosities are at times several orders of magnitude higher than those of BL Lacs (see Fig, panel), but whose masses are on average only a few times higher than the masses of BL Lacs (see Fig, panel), implying that FSRQs need higher Doppler factors to reach such luminosities. --Use in discussion, sec. 4.2

%The time-averaged spectra in the 157-month catalog are not up-to-date

The time-averaged spectra in the 157-month catalog have not yet been calculated, so for the comparison with the photon index $\Gamma$, we assume that the time-averaged spectra have not changed drastically over long timescales and therefore use the values for $\Gamma$ available in the previously published 105-month catalog\footnote{https://swift.gsfc.nasa.gov/results/bs105mon/}. The lack of correlation we observe here on monthly timescales is interesting since, naively, one might expect steeper indices to be indicative of radiative losses that may lead to significant flux variability. Our results suggest that, on these longer timescales, energetic losses may not be the origin of the variability. 

%Since for higher-energy particles, the cooling timescales are shorter, this would translate to higher and more rapid variability. Our results therefore suggest that, on these longer timescales, energetic losses may not be the origin of the variability.  --Use in discussion

%\begin{figure*}
%\centering
%    \includegraphics[width=0.47\textwidth]{0312broadbandspec.pdf}
%    \includegraphics[width=0.47\textwidth]{2MASSbroadbandspec.pdf}
%    \includegraphics[width=0.47\textwidth]{1RXSbroadbandspec.pdf}
%    \includegraphics[width=0.47\textwidth]{PKS2126abslogpar.pdf}
%\caption{Unfolded spectra and best-fit models for the broadband X-rays in our sample. The top two panels are fits with a simple power law.}
%\label{fig:bestfits}
%\end{figure*}

\begin{figure}
	% To include a figure from a file named example.*
	% Allowable file formats are eps or ps if compiling using latex
	% or pdf, png, jpg if compiling using pdflatex
	\includegraphics[width=\columnwidth]{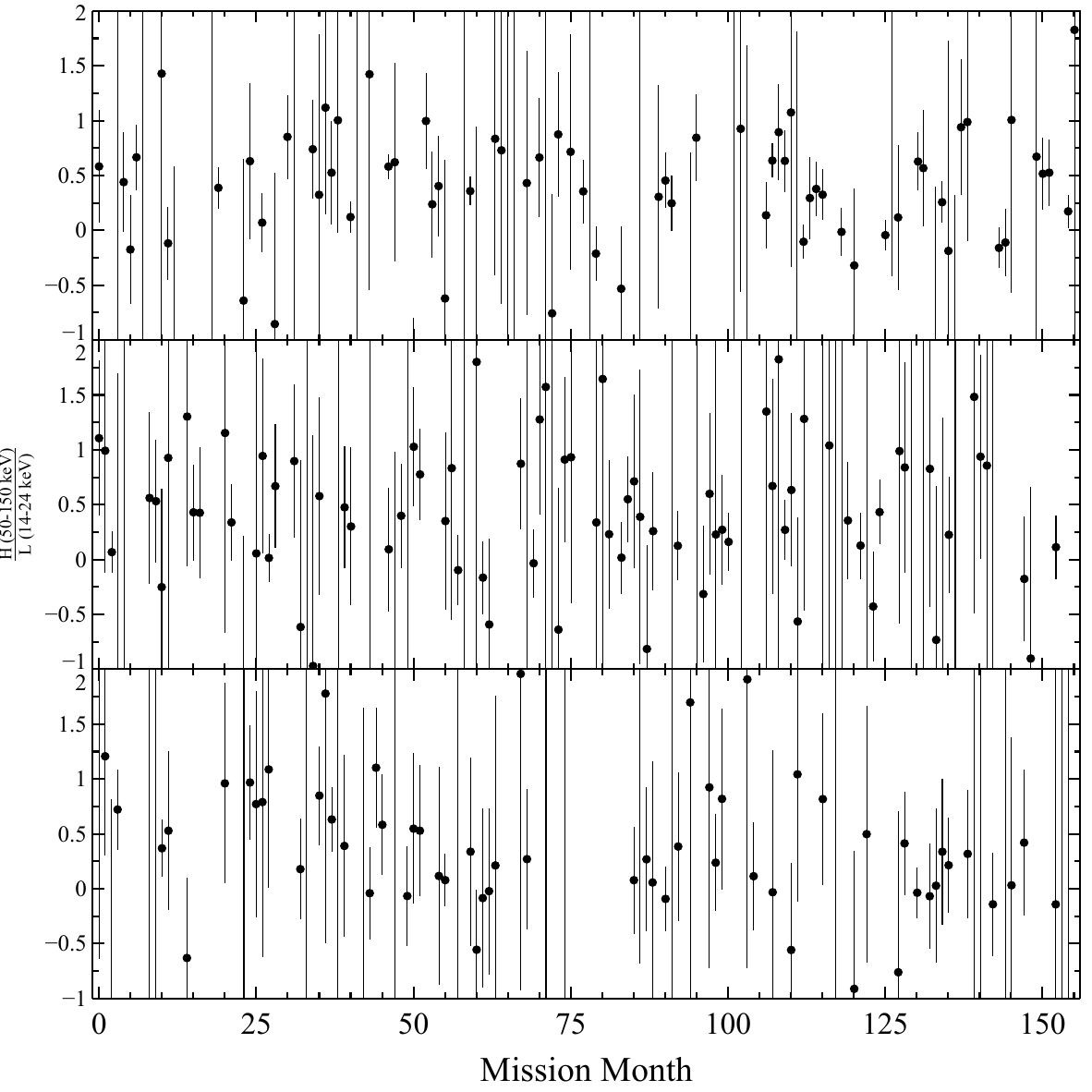}
    \caption{Most of the blazars in our sample do not show spectral variability on monthly timescales according to a $\chi^{2}$-test. Here, we show 3 blazars (top to bottom: 3C 279, 4C +04.42, B2 0743+25) for which this is the case, by showing the time series for the $\frac{H}{L}$ hardness ratio as an example. For the vast majority of the blazars, we do not detect statistically significant spectral variability due to the relatively low S/N of their eight-band light curve data; at times, there are also insufficient variations between the amplitudes of the points.}
    \label{fig:HLvstimenospecvar}
\end{figure}

\begin{figure*}
\centering
    \includegraphics[width=0.47\textwidth]{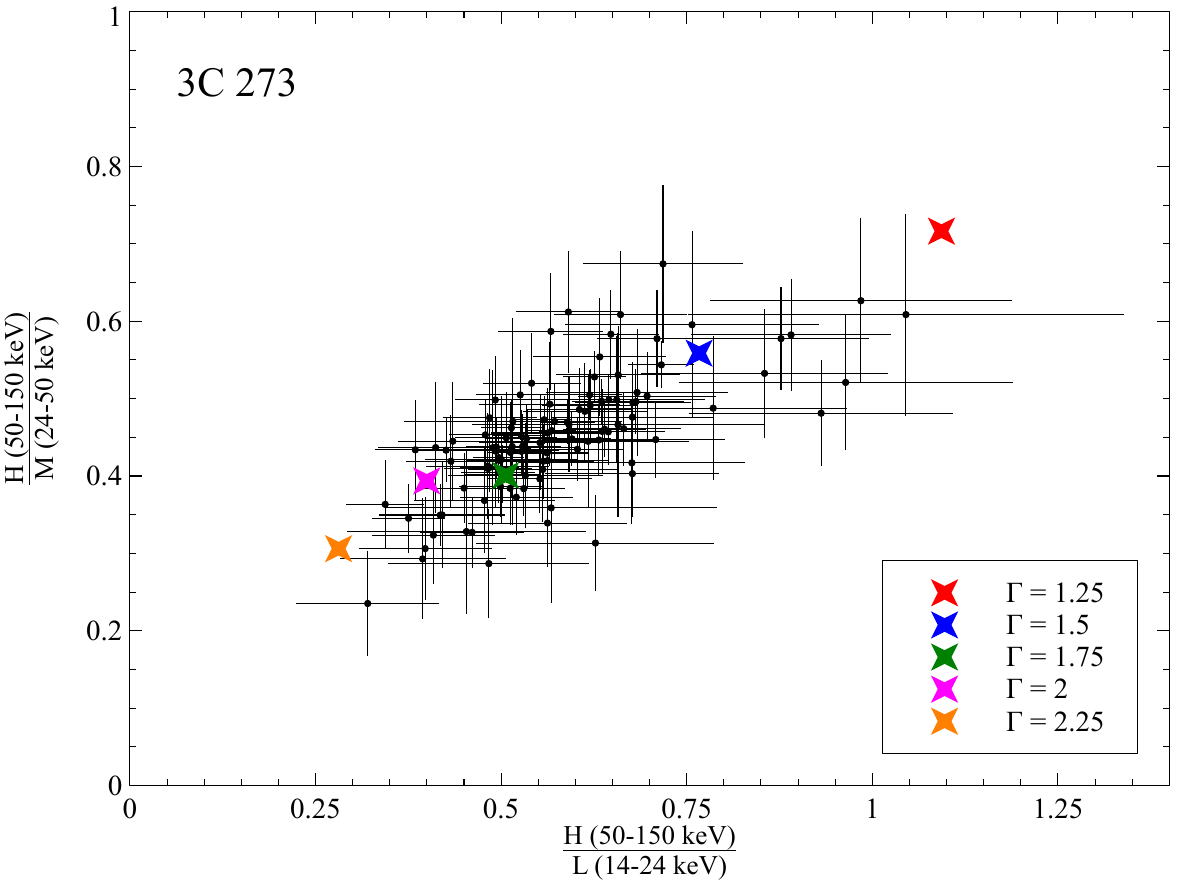}
    \includegraphics[width=0.47\textwidth]{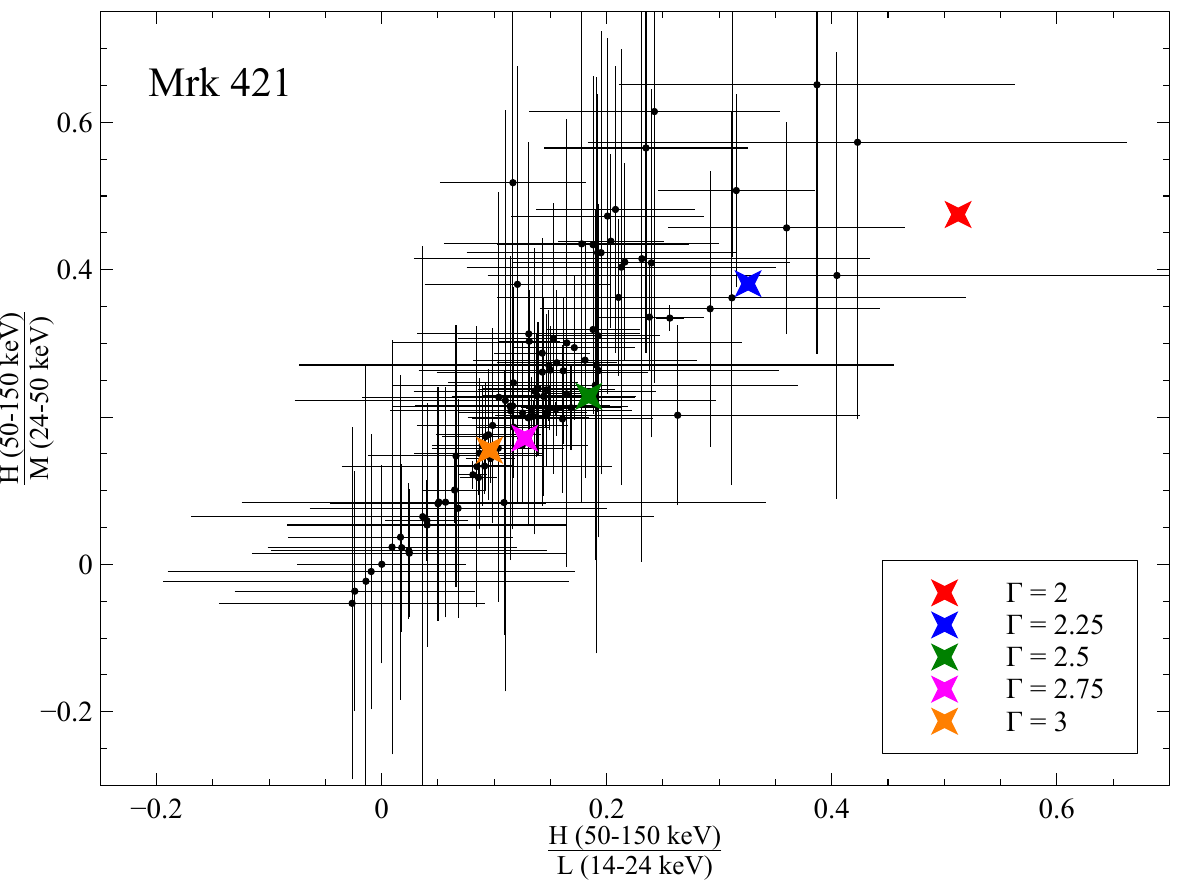}
    \includegraphics[width=0.47\textwidth]{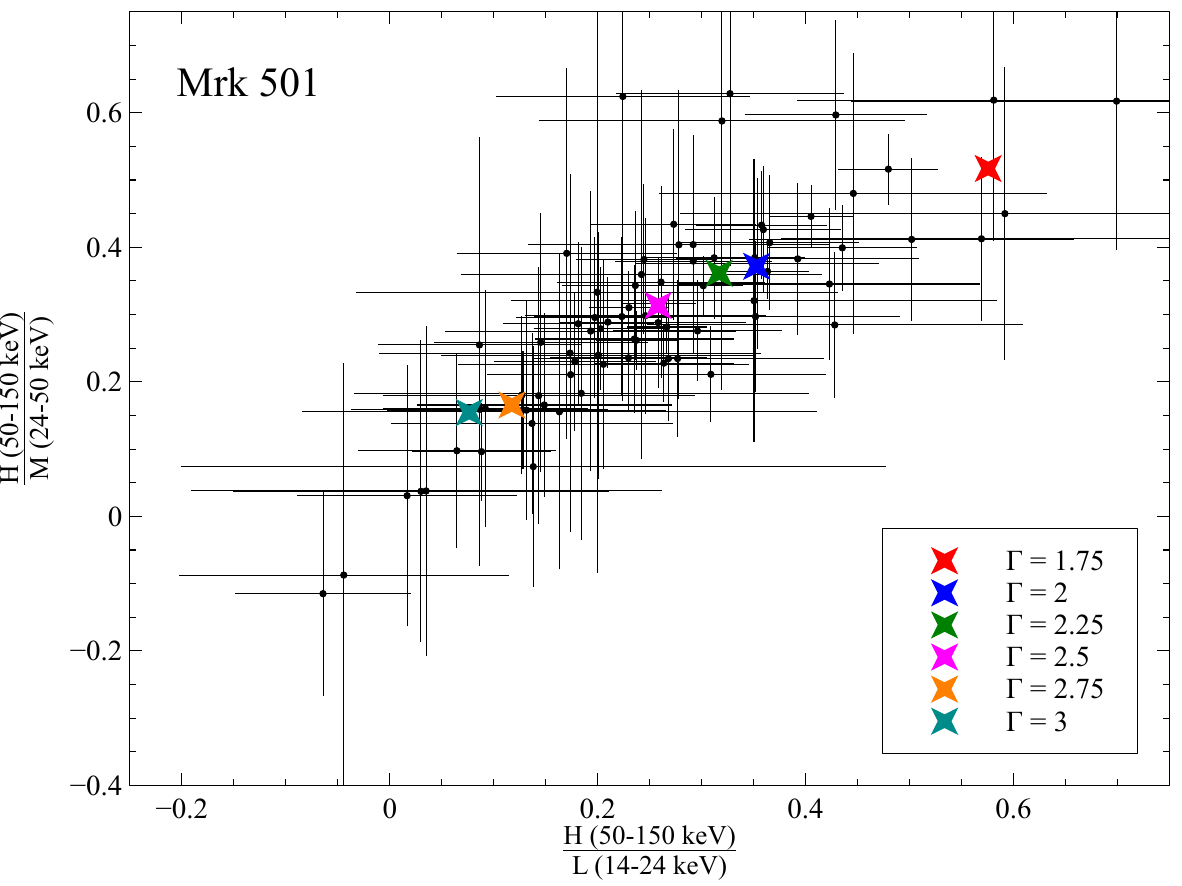}
\caption{HR planes for 3C 273, Mrk 421, and Mrk 501. The points calculated from the power law simulations (colored crosses) track the trend in the hardness ratios, suggesting that the spectra for these sources can be described with a simple power law that is variable on monthly timescales.}
\label{fig:HRplanes}
\end{figure*}

\subsection{Spectral variability analysis}

In order to detect potential variations in the spectra of our sources, we make use of the monthly light curves that are divided into the eight individual BAT bands. In addition to the filtering mentioned in Sec. 2, we also exclude months where the total-band count rate is negative to maximize the S/N for this particular analysis.

\subsubsection{Hardness ratio calculations}

We quantify the spectral variability by calculating different hardness ratio (HR) time series. To set up the HR calculations, for each month, we bin the eight BAT channels into three bins with similar average S/N over time. The reason for re-binning each monthly spectrum is two-fold: firstly, we would like to maximize the S/N before calculating the HRs. Secondly, having three bins would allow us to investigate the spectral variability in more detail, as it would allow for the calculation of at least three different simple hardness ratios. This results in the following channels: a ``low” (L) channel from 14-24 keV, a ``medium” (M) channel from 24-50 keV, and a ``high” (H) channel from 50-150 keV. We exclude the 150-195 keV channel due to its very low S/N. The relevant hardness ratios for this analysis are therefore $\frac{H}{L}$, $\frac{H}{M}$, and $\frac{M}{L}$.

As in the flux variability analysis, we fit each of the HR time series with a constant, and a $\chi^{2}$-test is applied, with the null hypothesis stating that the time series can be fit with a constant. Again, we define objects that show spectral variability are those for which $p_{\chi^{2}} < 5$\%. Upon applying this test, we find that only 5 sources (3 BL Lacs, 2 FSRQs) show spectral variability. This is once again in tension with the literature, which states that blazars on average can exhibit significant variability in their spectra; it is therefore surprising that we do not detect spectral variability for the vast majority of our sources in these long-term data. However, it is also possible that the BAT is simply not sensitive enough to detect such changes most of the time (see Figure \ref{fig:HLvstimenospecvar} for examples).

For the 5 sources that do show spectral variability (3C 273, 3C 454.3, Mrk 421, Mrk 501, and 1ES 0033+595), we extend the analysis further by attempting to characterize the nature of the changing spectrum on timescales of a month with a specific model. We decide to do this by plotting two of the calculated HRs against each other to form a ``HR plane" (see Figure \ref{fig:HRplanes} for examples); this essentially allows for an analysis of the monthly spectral data using only hardness ratio values, independent of time. Using the ``fakeit" command on XSPEC v12.12.0g (\citealt{1996ASPC..101...17A}), we then simulate monthly BAT spectra based on a simple power law, the simplest possible model for the X-ray emission of AGN. More specifically, we use the “pegpwrlw” model in XSPEC and the time-averaged 14-195 keV flux (in erg cm$^{-2}$ s$^{-1}$) of our sources to construct the simulated spectra. The simulations are performed with different power law slopes for each source, with $\Gamma$ ranging from $\sim$1 to 3, which is the typical range of photon indices of AGN in the BAT catalog. With these fake spectra, we can then calculate the HRs as we have done for the monthly data, and we plot the simulated HR values on the HR plane, shown as colored crosses in Figure \ref{fig:HRplanes}.

For 3 blazars, we notice a visual positive correlation between the $\frac{H}{L}$ and $\frac{H}{M}$ ratios from the data. We confirm this with a Pearson correlation analysis, with correlation coefficients of 0.70 for 3C 273 and 0.88 for both Mrk 421 and Mrk 501, all at a significance of $>$99.99\%. As shown in Figure \ref{fig:HRplanes}, the simulated points agree with the correlation in the data, suggesting a variable spectrum described by a power law with a photon index that changes on a month-to-month basis.

%In the past, the hard X-ray emission of 3C 273 has been described with a combination of 2 components: an ``AGN" component representing disk/coronal emission, and a jet component, with the implication that the spectral variability is a result of a change in the normalization of either component (see e.g. Madsen et al. 2015). In order to check for this, during the course of our analysis we did perform simulations for such a scenario, but found that the 2-component model is too complex given the quality of the BAT data, and that the hardness ratio values spanned by the data in the HR plane is best represented by the simulated HR points that result from the simple power law simulations. We therefore came to the ocnclusion that the higher sensitivity of NuSTAR is probably more appropriate to explore a complex scenario like a 2-component model.   --Use in discussion, sec. 4.3

We find that when using the HR plane method, we cannot conclusively describe the spectra of either 3C 454.3 or 1ES 0033+595 in the same way. We should note, however, that these sources are also different in that they show statistically significant variability for only two of the hardness ratios that we calculate, namely $\frac{M}{L}$ and $\frac{H}{L}$. The fact that these are the two ratios that vary suggests that these spectra may require at least a changing broken power law with a pivot at $\sim$25-30 keV in order to adequately describe their variability. However, the BAT data do not have a high enough S/N to be able to be described by such a complex model, as compared to a simple power law.

\subsubsection{Extracting values of $\Gamma$ from the hardness ratios}

\begin{figure}
	% To include a figure from a file named example.*
	% Allowable file formats are eps or ps if compiling using latex
	% or pdf, png, jpg if compiling using pdflatex
	\includegraphics[width=\columnwidth]{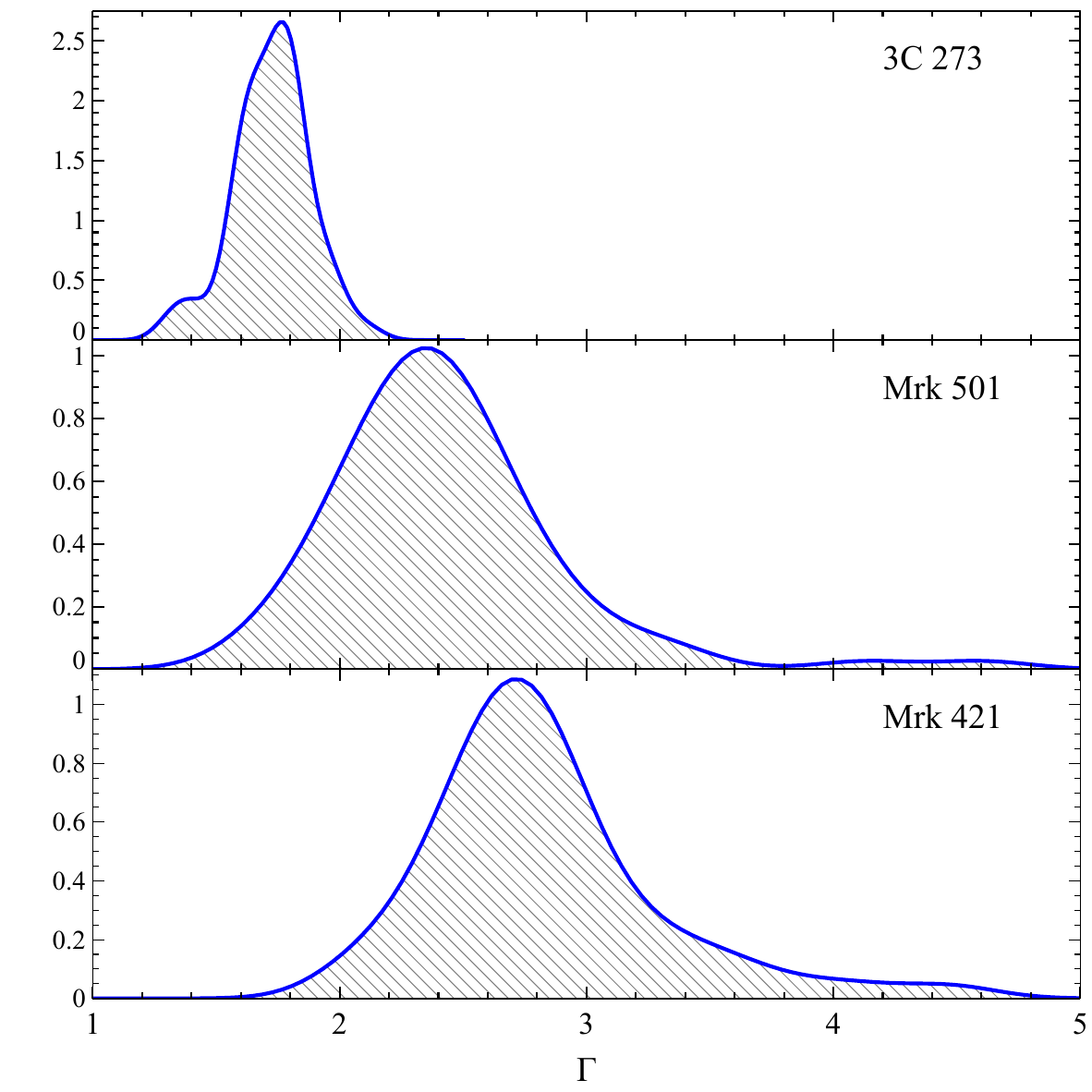}
    \caption{KDE distributions for $\Gamma$; the peaks of the distributions agree with the portions in the HR planes that are the most heavily populated with the data points As expected, 3C 273, an FSRQ, has overall a flatter spectrum, given that the X-rays are likely produced via the EC process, which has a hard spectral shape in the X-rays. By contrast, the X-rays in the two BL Lacs shown (both HSPs) are produced via synchrotron emission and lie on the tail of the synchrotron hump in the SED (see e.g. \citealt{2019ApJ...881..154P}).}
    \label{fig:PLdists}
\end{figure}

\begin{figure}
	% To include a figure from a file named example.*
	% Allowable file formats are eps or ps if compiling using latex
	% or pdf, png, jpg if compiling using pdflatex
	\includegraphics[width=\columnwidth]{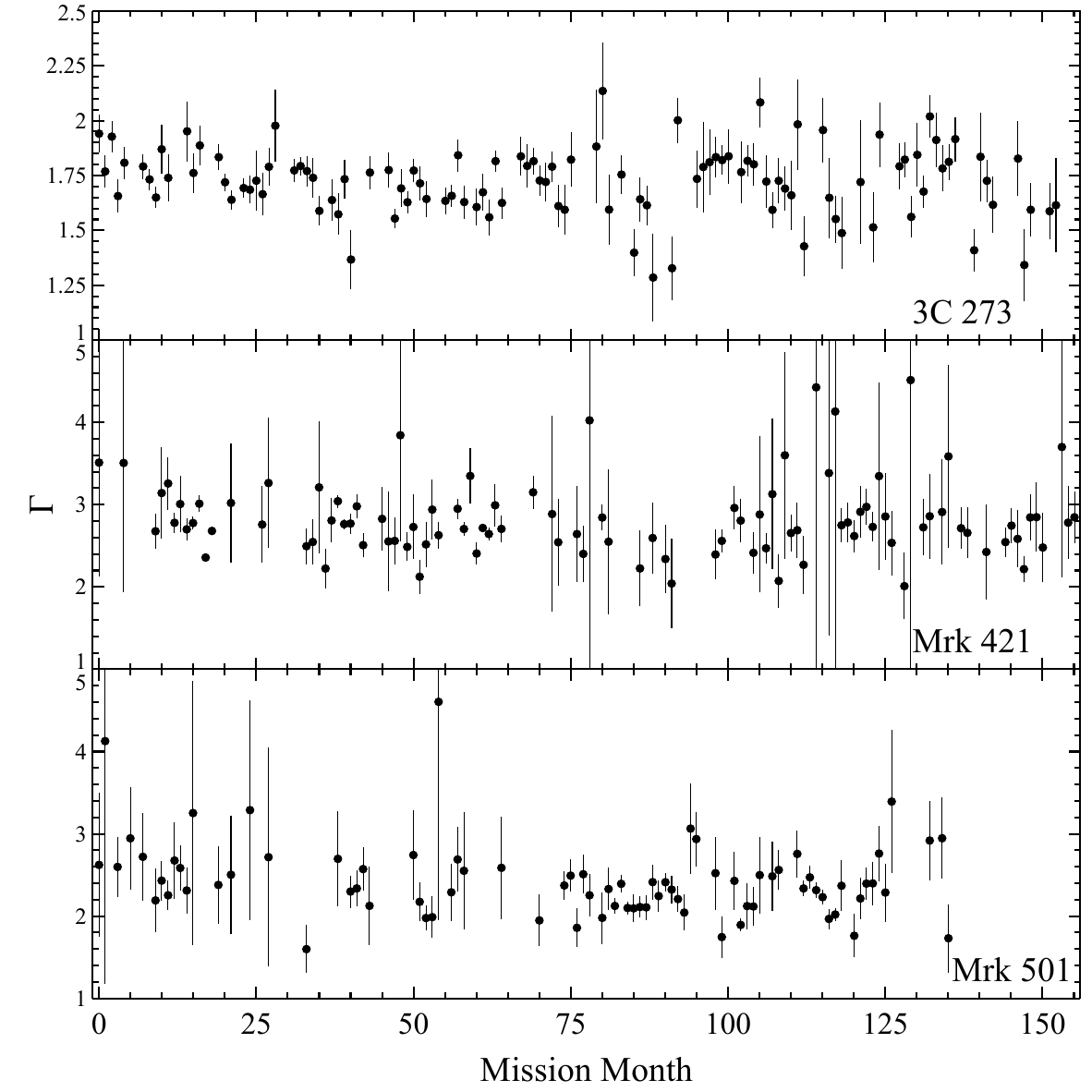}
    \caption{Photon index time series for 3C 273, Mrk 421, and Mrk 501. The values agree with the data in the HR plane plots, as well as the distributions; the data points missing at the end of the Mrk 501 light curve are due to the filtering of very low S/N points that have a total-band negative count rate.}
    \label{fig:PLseries}
\end{figure}

For the 3 sources whose spectra can be described by a power law, we push the spectral analysis even further. Since we have calculated the HRs for the data, as well as those for the simulated points based on a power law, we have the necessary tools to extract values for the photon index $\Gamma$ for each monthly data point from the monthly HR values. In order to do this, we plot either HR against the slopes used in the simulations, and find a relation for each source for the power law slope as a function of the HR. We show distributions for the values of the photon index for these 3 sources in Figure \ref{fig:PLdists}, as well as the photon index time series in Figure \ref{fig:PLseries}. It is clear from these plots and their mean values that the extracted values for $\Gamma$ agree with where in the HR plane the HRs cluster relative to the simulations. We also note the difference in the distributions between 3C 273 (an FSRQ) and Mrk 421, Mrk 501 (BL Lacs, both HSPs); this is consistent with the fact that FSRQs/LSPs generally have flatter spectra, due to the X-rays falling on the rising part of the high-energy hump in the SED (see e.g. \citealt{2017MNRAS.469..255G,2019ApJ...881..154P}, and Figure \ref{fig:corrs}).

\begin{figure*}
\centering
    \includegraphics[width=0.47\textwidth]{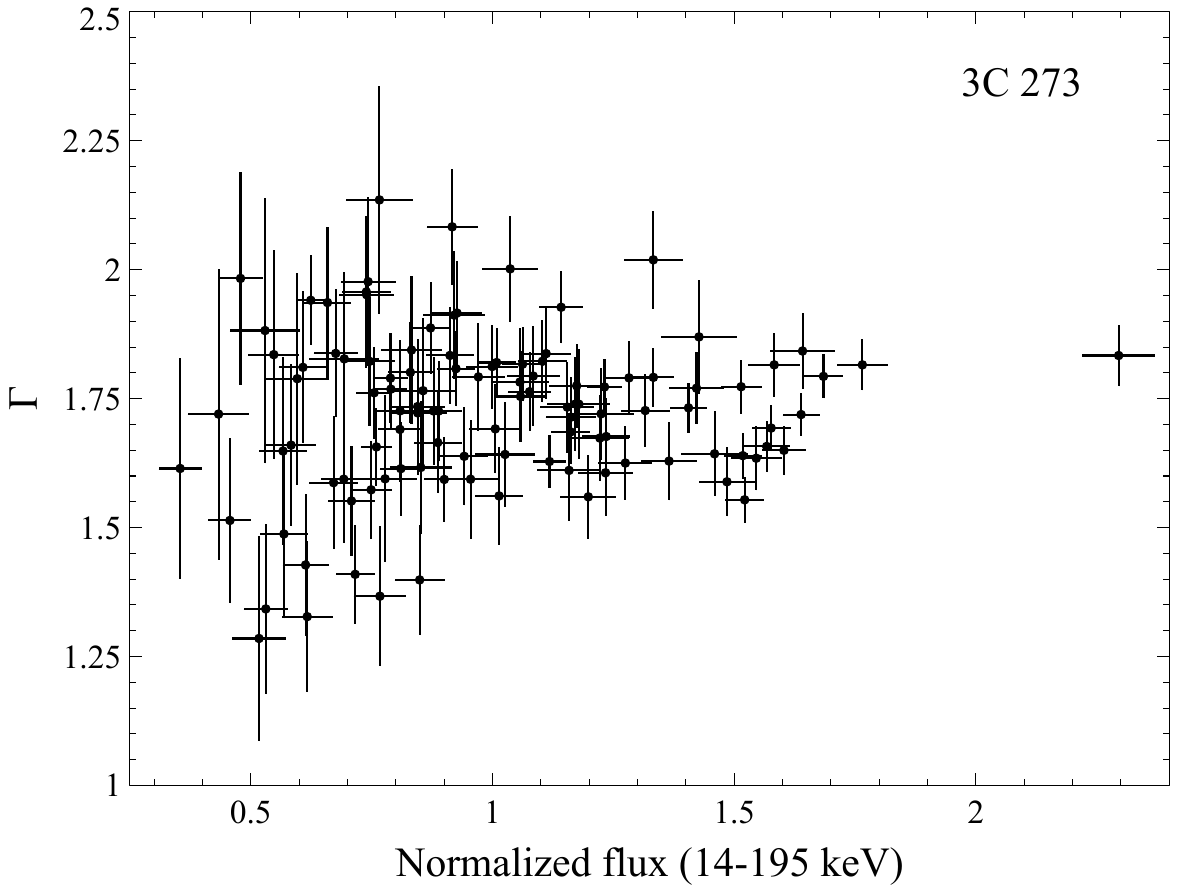}
    \includegraphics[width=0.47\textwidth]{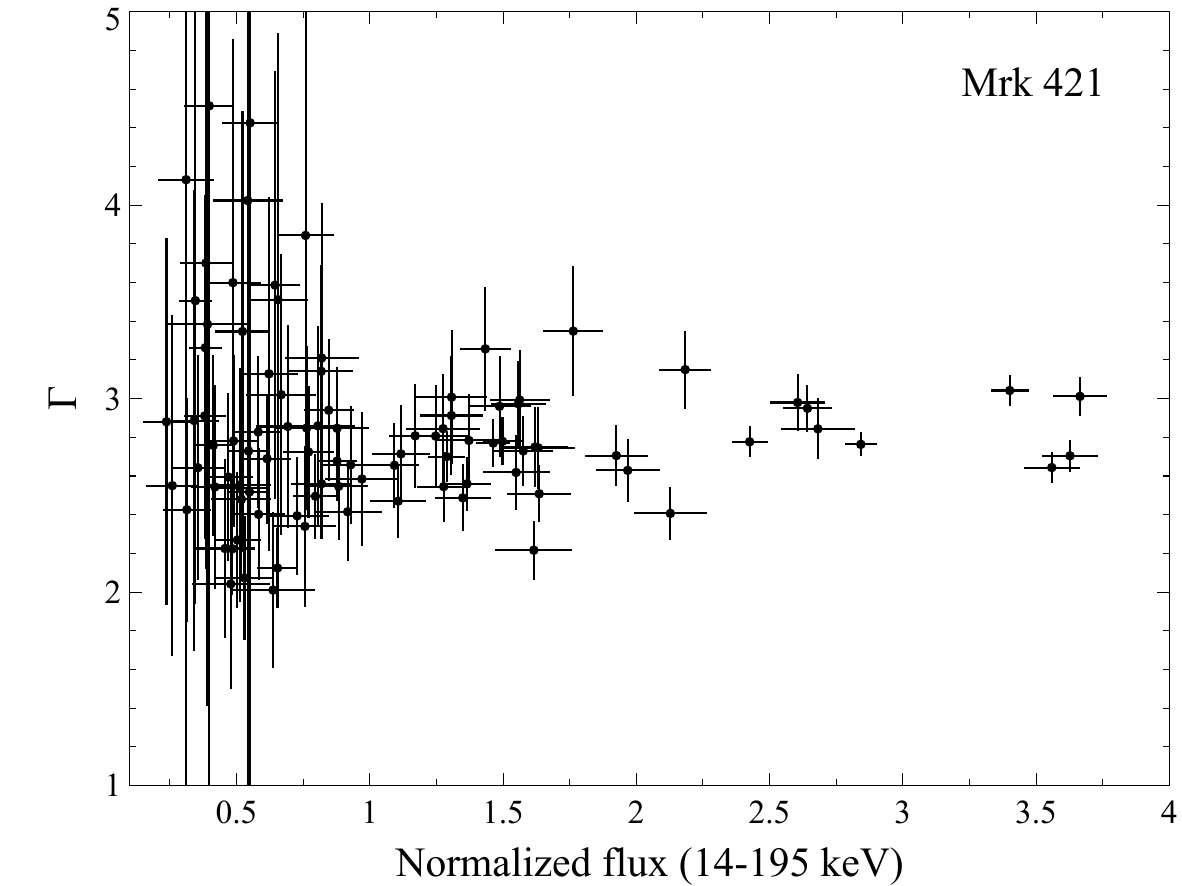}
    \includegraphics[width=0.47\textwidth]{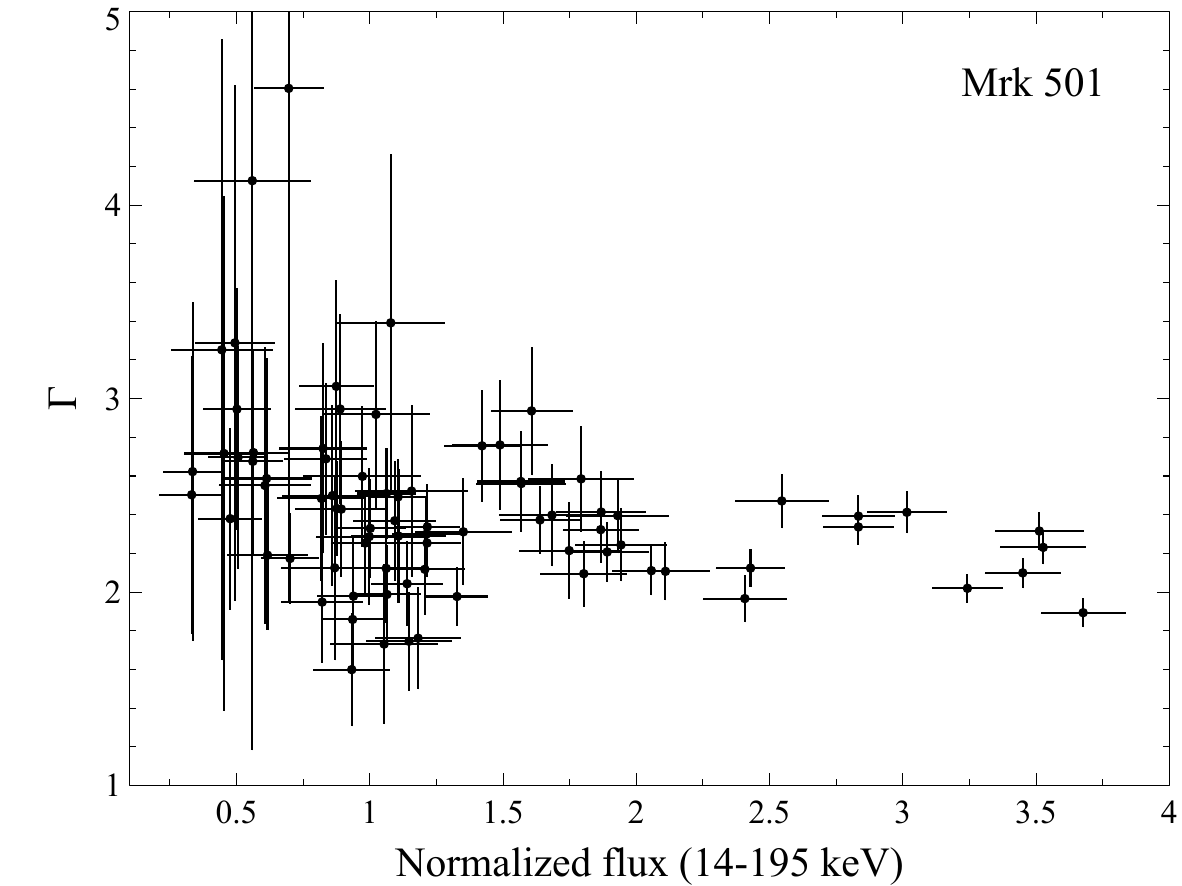}
\caption{$\Gamma$ as a function of the normalized flux for 3C 273, Mrk 421, and Mrk 501. We do not detect any correlation for 3C 273 or Mrk 421. The Mrk 501 data reveals a ``harder-when-brighter" trend, at a significance of $>$99.98\%.}
\label{fig:slopevsflux}
\end{figure*}

We also investigate how the spectra change with brightness, and we show the photon index as a function of the normalized 14-195 keV flux for these sources in Figure \ref{fig:slopevsflux}. We do not find a statistically significant trend for either 3C 273 or Mrk 421, but for Mrk 501, we detect a ``harder-when-brighter" behavior at a significance of $>$99.98\%, possibly associated with increased particle injection that hardens the particle energy distribution, resulting in a flatter emitted spectrum.

%then perform spectral simulations based on different power law slopes for each source (with the slopes based on the range of spectral slopes of objects in the BAT catalog), calculate the hardness ratios as we do for the data, and plot these simulated values on the hardness ratio plane     

\section{Discussion}\label{sec:sec4}

\subsection{Towards a more complete view of the X-ray variability of blazars}\label{sec:sec4.1}

%\subsubsection{Not your average blazar? Assessing a potential lack of variability}

Blazars have historically been described as extremely variable objects, showing significant variations in their flux and spectra across many timescales and wavelengths. However, it is well known that most X-ray studies of blazars have been biased towards the brightest sources, and/or towards blazars in active states such as flares. In this study, at least for a hard X-ray selected sample of sources, we attempt to reduce some that bias by analyzing the long-term data of the vast majority of the blazars in the \textit{Swift}-BAT 157-month catalog, which provides a more complete sample of blazars with varying degrees of brightness. 

By expanding the flux range covered in such an analysis, we find that it is possible that not all blazars are necessarily as highly variable as originally thought. However, the timescales on which variability is measured can be radically different in the literature, and simply stating that blazars are usually ``highly variable" must be qualified by the timescales covered. For this particular study, it is also important to stress that there are two major caveats to consider along with our seemingly surprising results. The first is that it is unclear from just the BAT data if this supposed lack of variability is due to real near-constant emission, or if it is solely due to sensitivity issues of the BAT data, given that the BAT does not have the highest sensitivity per unit time as compared to other X-ray telescopes. The second is that, on very long timescales, it is possible for blazars to exhibit fairly long periods of relative quiescence that are interrupted by flaring events. Indeed, studies such as the ones conducted by e.g. \cite{2014ApJ...789..135W}, \cite{2015ApJ...803...15P}, and \cite{2015ApJ...807...79H} suggest periods of relatively constant emission on close to yearly timescales in both the X-rays and other wavelengths, with some sources spending a significant amount of time in a quiescent state, implying mostly low-amplitude emission with occasional flaring events.

%before confidently making such a claim

%\begin{figure}
	% To include a figure from a file named example.*
	% Allowable file formats are eps or ps if compiling using latex
	% or pdf, png, jpg if compiling using pdflatex
%	\includegraphics[width=\columnwidth]{Lumhist.pdf}
%    \caption{bbloop papi plop}
%    \label{fig:Lumhist}
%\end{figure}

In \cite{2023MNRAS.520.1044M}, we investigate these potential scenarios for a sub-sample of 4 faint, ``quiescent" blazars that were taken from the population for which we did not detect variability that is mentioned in Section 3.1; the study involves the joint use of BAT data and data acquired from the \textit{NICER} observatory. \textit{NICER}'s $>$100 times sensitivity per unit time as compared to the BAT allowed us to probe shorter timescales that may not be detected by the BAT, while also allowing for an estimate of the variability on timescales similar to those of the BAT catalog. In general, the high sensitivity of the \textit{NICER} observatory therefore offered a way to possibly confirm the apparent lack of X-ray variability seen in the BAT data. In that study, we find that variability is in fact detected in the \textit{NICER} band, but that the variations are much lower amplitude than is expected of a blazar and appear to decrease with longer timescales. In addition, joint fits between the co-added \textit{NICER} spectra and the time-averaged BAT spectra suggest that any potential variability between the two bands would be occurring on timescales significantly longer than one year (see \citealt{2023MNRAS.520.1044M} for further details). Therefore, for at least 4 of the ``non-variable" blazars in this current study, it is possible that the X-ray emission is represented by periods of quiescence that are much longer than the ones typically observed in blazars.

The latter possibility, combined with the fact that for these 4 blazars, the amplitude of the variations decreases with increasing timescale, is an interesting result in that it is in tension with past studies that show that the stochastic variations in the emission of AGN can often be described with a ``red noise" power spectral density (PSD) that rises towards lower temporal frequencies, which would imply that the amplitude of the variability should instead increase with longer timescales (see e.g. \citealt{1999ApJ...514..682E,2002MNRAS.332..231U,2003ApJ...593...96M,2004MNRAS.348..783M,2005MNRAS.363..586U} for Seyfert-like AGN, and \citealt{2008ApJ...689...79C,2012ApJ...749..191C,2010ApJ...722..520A,2022ApJ...927..214G} for blazars). The results from \cite{2023MNRAS.520.1044M} on this sub-sample therefore suggest that, for the rest of the blazars here for which we do not detect variability in the 157-month BAT data on long timescales, it is at the very least possible that the power spectra have a different behavior on long timescales as compared to short timescales, and that they are possibly different from a red noise spectrum altogether.

For the 76 blazars for which we can detect statistically significant flux variability, we find that on average the FSRQs and BL Lacs in the sample have very similar $F_{\rm var}$. It is probably the case that our sample sizes are not large enough to confirm this as a general behavior, as we perform this analysis for only 53 FSRQs and 24 BL Lacs (the BAT blazar population in general is dominated by FSRQs). However, \cite{2020A&A...634A..80R} find similar results for the long-term $\gamma$-ray variability of a larger population of FSRQs and BL Lacs. They use data from the third \textit{Fermi}-LAT catalog to show that, for a significantly larger sample size, the variability of FSRQs is in fact significantly higher than that of BL Lacs. FSRQs produce the X-rays via EC effects, meaning that, assuming a leptonic model, external radiation arriving from extended regions like the BLR and dusty torus is inverse Compton scattered by the highly energetic electrons in the jet. In this scenario, due to light travel time effects, one might expect that the variability caused by variations in the external photon fields would be smeared out, leading to an observed overall lower-amplitude variability. Given the results here and in \cite{2020A&A...634A..80R}, it is therefore possible that the variability for some of these objects is instead driven by a variability in the injection function and in turn by changes in the particle energy distribution.

%which may not be detected by the BAT; was after "lower amplitude variability", changed the text

One physical scenario in which the above situation might arise involves a variability that is mostly caused by internal shocks moving along the jet. In this ``shock-in-jet" model (\citealt{1985ApJ...298..114M,2001MNRAS.325.1559S}), the injected energy is transmitted at irregular intervals to accelerate the shells of plasma in the jet. Energetic shocks in the jet then emerge via collisions between these shells, and some of the energy of the shocks is eventually converted into the radiative energy output of the relativistic particles, resulting in both a variable particle energy distribution and variable emission. Our results suggest that this may be a significant contributor to the variability not just in BL Lacs (where it is expected that the variability involves a balance between the acceleration and synchrotron/SSC cooling timescales of the relativistic electrons in the jet), but also in FSRQs.

\subsection{Relationships between $F_{\rm var}$ and important blazar parameters}

Several groundbreaking studies have shown that the fractional variability of AGN is anti-correlated with the sources' luminosity and black hole mass (\citealt{1986Natur.320..421B,1993MNRAS.265..664G,1995MNRAS.273..923P,2001MNRAS.324..653L,2002MNRAS.332..231U,2004MNRAS.348..207P,2010ApJ...710...16Z,2011ApJ...730...52K,2013ApJ...779..187K,2012A&A...542A..83P}). These results have usually been interpreted in the context of accretion timescales and are thus also linked to additional correlations between e.g. the break frequency in the PSD of these sources and the black hole mass and accretion rate (\citealt{2003ApJ...593...96M,2006Natur.444..730M,2007MNRAS.380..301K,2011ApJ...730...52K}); the break frequency $\nu_{\rm b}$ in particular has come to be recognized as a vital quantity to describe accretion processes and specific timescales of accretion. However, these studies analyzed the variability usually only in the 2-10 keV band, finding that $\nu_{\rm b}$ occurred on relatively short timescales.

\cite{2013ApJ...770...60S} performed one of the first PSD analyses with BAT data using the 58-month catalog for 30 AGN. Due to the long timescales probed, they could not detect breaks in the power spectra, with the ensuing implication being that there was a lack of correlation between the variability and the luminosity and black hole mass. \cite{2014A&A...563A..57S} similarly ascribe the same lack of correlation in their analysis to the fact that, for the vast majority of the sources in the catalog, the BAT can only constrain the variability on monthly timescales, which is much longer than the typical PSD break times that usually dictate some of these correlations in AGN. 

In this study, we do not find significant correlations between the fractional variability of blazars on monthly timescales and their luminosity and black hole mass. Since we probe the same long timescales as in the two aforementioned studies, it is possible that our results arise for similar reasons. However, one major difference is that we focus on blazars, whereas the previous studies are exclusive to Seyferts. Even though, for some blazars, a correlation between variability and black hole mass could have led to a superficial probing of the disk-jet connection, in general, the blazars in our sample have very different underlying astrophysical processes associated with their X-ray emission (jetted emission) compared to Seyferts (emission from the disk and corona). Therefore, it is unclear whether this comparison with past studies is 100\% valid to begin with.

While we also do not find a correlation between $F_{\rm var}$ and the Doppler factor $\delta$, we do observe that the FSRQs have a higher mean $\delta$ than BL Lacs. This might be expected of the more luminous FSRQs, whose luminosities are at times several orders of magnitude higher than those of BL Lacs, but whose black hole masses are on average only a few times higher than those of BL Lacs (see Figure \ref{fig:corrs}), implying that FSRQs need higher Doppler factors to reach such luminosities and that relativistic beaming may be contributing significantly to the X-ray variability in these sources.

The lack of correlation between $F_{\rm var}$ and the photon index $\Gamma$ is somewhat unexpected, since the steeper indices associated with a significantly falling spectrum are likely indicative of radiative losses. Since for higher-energy particles, the cooling timescales are shorter ($t_{\rm cool} \propto \gamma^{-1}$), this would translate to higher and more rapid variability. In a study with \textit{NuSTAR} data, \cite{2018A&A...619A..93B} do find a positive correlation between the hard X-ray variability and photon index (for both FSRQs and BL Lacs separately) for several observations of a sample of 13 blazars. They find that for BL Lacs, the trend is not as distinct as is found for FSRQs; they attribute this to rapid synchrotron cooling in BL Lacs at higher energies possibly contributing to photons at the high-energy end of each individual spectrum, resulting in a harder power law distribution and thus reducing the relationship seen of a higher variability with steeper index. This study, however, is performed on the shorter timescales probed by \textit{NuSTAR}, and they note that their sample size is limited. The fact that we do not see any correlation in our study therefore suggests that cooling processes are not necessarily at the heart of the variability on longer timescales.       

%Our results therefore suggest that, on these longer timescales, energetic losses may not be the origin of the variability

%Eveen though, for some sources, a correlation between variability and black hole mass could have led to a superficial probing of the disk-jet connection, in general, the blazars in our sample have very different underlying astrophysical processes associated with their X-ray emission (jetted emission) compared to Seyferts (emission from the disk and corona). Therefore, it is unclear whether this comparison with past studies is valid to begin with. 

\subsection{Interpreting the spectral variability}

We detect statistically significant variability in 5 of our sources. For 3 of these, we find that the spectrum can be described with a simple power law that changes spectral slope on monthly timescales. This is generally in line with studies that show that the X-ray spectra (and in particular, the hard X-ray spectra) of blazars can be described with a power law. However, there are hints in our analysis pointing to potentially more complex spectra. For example, for 3C 454.3 and 1ES 0033+595, only two hardness ratios, $\frac{M}{L}$ and $\frac{H}{L}$, show statistically significant variability. This suggests that for these sources, the hard X-ray spectra may require a broken power law model to adequately describe their variability, with spectral variations possibly arising from the BAT spectra pivoting about a break energy at $\sim$25-30 keV. Such a break in the spectra could be present due to the BAT data lying between the synchrotron and Compton humps, as usually the hard X-ray data on either hump have significantly different spectral slopes (lower $\Gamma$ for inverse Compton hump, higher $\Gamma$ for synchrotron hump); alternatively, curvature could be detected if the data lie on the very peak of either hump. We inspect the latest SEDs provided for these sources in \cite{2019ApJ...881..154P}, but find that all of the BAT data for 1ES 0033+595 are on the tail of the synchrotron hump, while all of the BAT data for 3C 454.3 are on the rising part of the Compton hump. However, given the significant variability we detect, it is possible that the SEDs shown in \cite{2019ApJ...881..154P}, and therefore the location of the X-rays in the SEDs, experience significant changes over long timescales. In an analysis of the long-term variability of 1ES 0033+595, \cite{2022A&A...668A..75K} do in fact find significant curvature in most of the source's \textit{Swift}-XRT spectra in the 0.3-10 keV band, stating that their best-fit model is a log-parabolic power law. Therefore, it is possible that our results are consistent with their analysis in that 1ES 0033+595 may require a more complex model than a simple power law.

The discussion of spectra more complex than a power law naturally brings up one of the most famous and well-studied sources in our sample. 3C 273 happens to be a blazar-like AGN (usually classified as an FSRQ within the context of blazars) with a fairly high viewing angle of $\sim$10$^{\circ}$ (\citealt{2004ApJ...613..119S}). While our analysis suggests that its emission can be described by a simple hard power law, several studies have indicated that its X-ray spectrum can be modeled by the combination of a Seyfert-like component (e.g. coronal emission and X-ray reflection features) and a beamed, blazar-like component to describe emission from the jet (\citealt{2002MNRAS.336..932K,2008A&A...486..411S,2015A&A...576A.122E,2015ApJ...812...14M}). In particular, the analysis in \cite{2015ApJ...812...14M} with \textit{NuSTAR} and \textit{INTEGRAL} data suggests that the hard X-ray spectral variability is caused by a change in amplitude in each component as their respective power law slopes remain fairly invariant. In an attempt to find some agreement through the BAT data, we briefly investigate this by performing simulations for such a scenario, but find that the quality of the BAT data is not high enough to detect the spectral features associated with the 2-component model, and that the hardness ratio values spanned by the data in the HR plane are best represented by the simulated HR points that result from the simple power law simulations. We therefore come to the conclusion that the higher sensitivity of \textit{NuSTAR} is probably more appropriate to explore a more complex scenario like a 2-component model.

Finally, we find that Mrk 501 shows signs of ``harder-when-brighter" behavior on monthly timescales, suggesting that the hard X-ray spectrum flattens as the source reaches brighter states. \cite{2018A&A...619A..93B} also find such a correlation for Mrk 501 with their $\textit{NuSTAR}$ analysis on shorter timescales. A reason for this could be that the brightening of the source corresponds to increased particle injection. As a HSP BL Lac, the BAT data for Mrk 501 lie at the tail of the synchrotron hump; this could cause the particle distribution to harden, resulting in a harder power law slope for the emitted spectrum. Alternatively, a strengthening of the magnetic fields in the jets could also lead to increased synchrotron emission, possibly producing harder photons in the process.  

%Since NuSTAR overlaps in energy with the BAT (at 14-75 keV), it would be interesting to use the higher resolution of NuSTAR to further investigate this and test whether that data requires a break in the spectrum.    

%statistically significant variability for only two of the hardness ratios that we calculate, namely $\frac{M}{L}$ and $\frac{H}{L}$. The fact that these are the two ratios that vary suggests that this spectrum may require at least a changing broken power law with a pivot at $\sim$25-30 keV in order to adequately describe its variability.

%(e.g. Arcodia et al. 2018, Dalton et al. 2021).    

%Therefore, one way to interpret the significant absorption that we observe in the soft X-rays is to infer that there is a possible contribution from the IGM. Absorption in the IGM would not change over time, given that it should not depend on the source's environment. In addition, the IGM is diffuse and smeared over redshift, meaning that it is possible for its signature to also appear at or near the redshift of the source.

%obtained in quiescent and active states in the optical, X-rays, and $\gamma$-rays.Williamson et al. (2017) also conducted a multi-wavelength study on a sample 33 blazars that divides the data into quiescent and active states, with the intent of comparing the spectral indices obtained in quiescent and active states in the optical, X-rays, and $\gamma$-rays.     

\section{Conclusions}

We have presented long-timescale, time-domain variability analyses of 127 blazars from the \textit{Swift}-BAT 157-month catalog by use of $\sim$13 years of continuous archival hard X-ray data in the 14-195 keV band. Our main results are as follows:

\begin{enumerate}
    \item We do not detect statistically significant flux variability for a significant fraction ($\sim$37\%) of the blazars in our sample, which is in tension with the expected highly variable emission of blazars on most timescales and wavelengths.   
    \item On average, for the objects that do show variability, we find that FSRQs appear to have a very similar degree of flux variability compared to BL Lacs ($\langle F_{\rm var} \rangle = 76\pm5$\% vs $\langle F_{\rm var} \rangle = 75\pm4$\%), possibly due to the variability in FSRQs being driven by variations in the particle injection as opposed to variations in external radiation fields. 
    \item We do not find correlations between $F_{\rm var}$ and the luminosity and black hole mass, possibly due to the fact that the BAT probes timescales much longer than the timescales where these correlations have previously been observed; however, the physical mechanisms producing the X-rays in these blazars are also significantly different to those in the Seyferts analyzed in these previous studies. We also do not find a trend between $F_{\rm var}$ and the photon index $\Gamma$, suggesting that radiative losses may not be the main source of variability for these objects on long timescales.
    \item We detect spectral variability in 5 blazars, and for 3 of them the behavior can be summarized as a simple power law in the hard X-rays that changes spectral slope $\Gamma$ on monthly timescales. For at least two sources, it is possible that a more complex model is required to describe the variable spectra. 
    \item For Mrk 501, a HSP BL Lac, we detect a ``harder-when-brighter" behavior at a significance of $>$99.98\%, possibly associated with increased particle injection or an enhancement of the magnetic fields in the jet.   
\end{enumerate}

\section*{Data Availability}

Supplementary data such as the entirety of Table \ref{tab:blazarprops} is available in the electronic version or upon request (contact Sergio A. Mundo). The archival \textit{Swift}-BAT data for the blazars are available at \href{https://swift.gsfc.nasa.gov/results/bs157mon/}{https://swift.gsfc.nasa.gov/results/bs157mon/}.

%%%%%%%%%%%%%%%%%%%% REFERENCES %%%%%%%%%%%%%%%%%%

% The best way to enter references is to use BibTeX:

\bibliographystyle{mnras}
\bibliography{example} % if your bibtex file is called example.bib

% Alternatively you could enter them by hand, like this:
% This method is tedious and prone to error if you have lots of references
%\begin{thebibliography}{99}
%\bibitem[\protect\citeauthoryear{Author}{2012}]{Author2012}
%Author A.~N., 2013, Journal of Improbable Astronomy, 1, 1
%\bibitem[\protect\citeauthoryear{Others}{2013}]{Others2013}
%Others S., 2012, Journal of Interesting Stuff, 17, 198
%\end{thebibliography}

%%%%%%%%%%%%%%%%%%%%%%%%%%%%%%%%%%%%%%%%%%%%%%%%%%

%%%%%%%%%%%%%%%%% APPENDICES %%%%%%%%%%%%%%%%%%%%%

%\appendix

%\section{Some extra material}

%If you want to present additional material which would interrupt the flow of the main paper,
%it can be placed in an Appendix which appears after the list of references.

%%%%%%%%%%%%%%%%%%%%%%%%%%%%%%%%%%%%%%%%%%%%%%%%%%

% Don't change these lines
\bsp	% typesetting comment
\label{lastpage}
\end{document}